\documentclass[superscriptaddress,a4paper,tightenlines,nofootinbib,floatfix,11pt]{article}
\usepackage{jheppub}

\usepackage{hyperref}
\usepackage{url}
\usepackage{graphics}
\usepackage{amsmath}
\usepackage{cancel}
\usepackage{bm}
\usepackage{graphicx}
\usepackage{subcaption}
\usepackage{booktabs}
\usepackage{siunitx}
\usepackage{amssymb}
\usepackage{relsize}
\usepackage{physics}
\usepackage{chemarrow}
\usepackage{multirow}
\usepackage{soul}
\usepackage[normalem]{ulem}

\DeclareUnicodeCharacter{2212}{-}

\newcommand{\chonepm}{\tilde{\chi}_1^{\pm}}

\def \st1{{\widetilde{t_1}}}
\def \mst1{m_{\widetilde{t_1}}}

\def \lspone{\tilde\chi_1^0}
\def \mlspone{m_{\lspone}}
\def \lsptwo{\tilde\chi_2^0}
\def \mlsptwo{m_{\lsptwo}}

\def\chonepm{\tilde{\chi}_1^{\pm}}
\def\chonemp{\tilde{\chi}_1^{\mp}}
\def\mchonepm{m_{\chonepm}}
\def\champ2{\tilde{\chi}_2^{\mp}}

\def \met{\rm E{\!\!\!/}_T}

\title{Reconstructing Sparticle masses at the LHC using Generative Machine Learning}

\author[1]{Rahool Kumar Barman,}
\author[2]{Arghya Choudhury}
\author[2]{Subhadeep Sarkar}
\affiliation[1]{Kavli IPMU (WPI), UTIAS, The University of Tokyo, Kashiwa, Chiba 277-8583, Japan}
\affiliation[2]{Department of Physics, Indian Institute of Technology Patna, Bihar - 801106, India}

\emailAdd{rahool.barman@ipmu.jp}
\emailAdd{arghya@iitp.ac.in}
\emailAdd{subhadeep\_1921ph21@iitp.ac.in}

\abstract{We explore a generative-model framework to infer the masses of heavy particles from detector-level data over a broad parameter space. Our model combines a transformer-based detector encoder and a diffusion neural network.  We first apply our model to a new physics scenario involving the pair production of wino-like chargino-neutralino, $pp \to \chonepm \lsptwo$, in the $1\ell + 2\gamma + jets$ channel at the high luminosity LHC~(HL-LHC). We find that our framework can achieve mass reconstruction efficiency of $\gtrsim 70\%$ for the lightest neutralino $\lspone$ and $\gtrsim 40\%$ for the second lightest neutralino $\lsptwo$, for a mass tolerance of $\Delta m = 30~$GeV, across the entire parameter space accessible at the HL-LHC. We further extend our analysis to a different scenario with $pp\to\chonepm\chonemp+\chonepm\lsptwo$ pair production at the HL-LHC in the $4\ell+\cancel{E}_T$ channel, and for a fixed value of $m_{\lsptwo}$, we obtain reconstruction efficiencies $\gtrsim80\%$ over a wide range of $\mlspone$ for $\Delta m = 30~$GeV.}

\begin{document}

\maketitle
\section{Introduction}
\label{sec:intro}

Despite the remarkable success of the Standard Model (SM), it has several shortcomings, such as the hierarchy problem \cite{SUSSKIND1984181,PhysRevD.14.1667}, origin of neutrino mass \cite{KamLAND:2013rgu,Borexino:2013zhu,RENO:2018dro,T2K:2018rhz,DayaBay:2018yms,Super-Kamiokande:2019gzr,NOvA:2019cyt}, the nature of dark matter (DM) \cite{Zwicky:1933gu,1937ApJ....86..217Z,Jungman:1995df,Sofue:2000jx}, etc. Many beyond the Standard Model (BSM) frameworks have been proposed, but we are yet to observe a definitive signature of new physics. As we enter the precision era of the Large Hadron Collider (LHC), with the upcoming high-luminosity LHC~(HL-LHC) set to collect roughly an order of magnitude more data than the current runs, we will be able to enhance sensitivity to several currently `rare' processes and parameterize potential deviations from the SM with far greater accuracy. However, to fully leverage the statistical boost, there is a need to develop improved analysis techniques capable of performing new physics inference, event reconstruction, and particle identification, with greater precision and speed.

Generative Machine Learning~(ML) techniques have shown an impressive potential to tackle these challenges, through their ability to perform high-dimensional event unfolding, fast event simulation, and event likelihood estimation directly from measured data~\cite{deOliveira:2017pjk,Paganini:2017hrr,Paganini:2017dwg,Datta:2018mwd,Butter:2019cae,Bellagente:2019uyp,Belayneh:2019vyx,Gao:2020zvv,Carminati_2020,Bellagente:2020piv,Butter:2020tvl,Krause:2021ilc,Kansal:2021cqp,Bieringer:2022cbs,Touranakou:2022qrp,Butter:2022vkj,ATLAS:2022jhk,Backes:2022sph,Mikuni:2023dvk,Ackerschott:2023nax,Butter:2023ira,Butter:2023llt,Huetsch:2024quz,Butter:2024vbx,Abasov:2025ntj,Chatterjee:2025gej}. For example, Generative Adversarial Networks~\cite{goodfellow2014generative} have been explored to perform fast calorimeter shower generation, detector effect simulation~\cite{deOliveira:2017pjk,Paganini:2017hrr,Paganini:2017dwg} and event unfolding~\cite{Datta:2018mwd,Butter:2019cae,Bellagente:2019uyp}, while Variational Autoencoders~\cite{rezende2014,kingma2022} have been used to perform fast simulation of jets at the LHC~\cite{Touranakou:2022qrp}. Normalizing Flows~\cite{dinh2015nice} have also demonstrated the ability to perform high-dimensional unfolding and full reconstruction of the parton-level phase space from detector-level data in a probabilistically faithful way~\cite{Bellagente:2020piv,Ackerschott:2023nax,Butter:2023llt,Abasov:2025ntj}, as well as fast-detector simulations~\cite{Krause:2021ilc}. More recently, diffusion models~\cite{sohldickstein2015deep}, have shown the ability to perform accurate reconstruction of high-dimensional target densities, with stable training dynamics~\cite{Mikuni:2023dvk,Butter:2023ira}. All together, these deep generative techniques show a promising landscape to extend both the precision and discovery reach of BSM searches at the HL-LHC.

In the present study, we focus on addressing a complementary question: if the HL-LHC observes an excess in a particular channel, how accurately can we reconstruct the mass spectrum of the underlying new states from detector-level information? As a test scenario, we consider a R-parity violating (RPV) SUSY framework\footnote{For some recent phenomenological works on RPV SUSY model, see references \cite{Dreiner:2023bvs,Choudhury:2023lbp,Choudhury:2023eje,Choudhury:2023yfg,Choudhury:2024yxd,Baruah:2024wrn,Dreiner:2025kfd} } \cite{Dreiner:1997uz,Martin:1997ns,Barbier:2004ez,Choudhury:2024ggy} with only non-zero $\lambda_{112}^{\prime\prime}$ UDD type coupling \footnote{To demonstrate the model independence of our ML network, we consider a different RPV scenario which involves lepton number violating LLE-type $\lambda_{121}$ and/or $\lambda_{122}$ coupling (See Appendix A)}. We analyze the wino-like electroweakino production at the HL-LHC, $pp\to\chonepm\lsptwo$, followed by the decays $\chonepm \to W^{\pm} \lspone$ and $\lsptwo \to h \lspone$. The bino-like lightest neutralino ($\lspone$) decays to $uds$ via $\lambda_{112}^{\prime\prime}$ coupling.
Additionally, the $W$ originated from chargino ($\chonepm$) decays leptonically ($W\to\ell^{\prime}\nu_{\ell^{\prime}}$, $\ell^{\prime}\equiv e,\mu,\tau$) and the higgs from $\lsptwo$ decays to a pair of photons ($h\to\gamma\gamma$). This leads to a final state, 
\begin{equation}
    pp \to \chonepm\lsptwo \to (\chonepm \to W^{\pm} \lspone) (\lsptwo \to h \lspone) \to (W^{\pm} \to \ell^\prime \nu) (h \to \gamma\gamma) + jets. 
    \label{eq:eqn:signal_proc}
\end{equation}
We choose the di-photon final state as it offers a clean signature despite a smaller branching fraction. We illustrate a representative Feynman diagram for our signal process in Fig.~\ref{fig:feynman_diag}. 
\begin{figure}[!t]
\begin{center}
\includegraphics[width=0.5\textwidth]{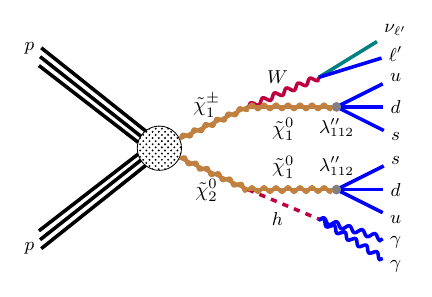}
\caption{Feynman diagram for SUSY signal where mass degenerate pair produced $\chonepm\lsptwo$ undergo cascade decays to $\lspone$ followed by decay to $uds$ via $\lambda_{112}^{\prime\prime}$ coupling. The $W$ boson decays to a lepton ($\ell^{\prime}\equiv e,\mu,\tau$) and one neutrino ($\nu_{\ell^{\prime}}$), and higgs boson ($h$) decays to two photons ($\gamma$).}
\end{center}
\end{figure}
\label{fig:feynman_diag}

Our goal is to learn a conditional mapping between detector-level kinematics and the parton-level four-momentum of electroweakinos, such that a single network can perform mass reconstructions over a wide parameter space. Since the $\lsptwo$ and $\chonepm$ masses are degenerate within the simplified scenario considered here, we restrict ourselves to reconstructing only the four-momentum for $\lsptwo$ and its daughter $\lspone$, keeping the setup as simple as possible. 


To this end, we explore a neural network architecture that combines two components. The first component is a transformer-based `detector encoder' that processes the per-particle four-momentum, along with a small set of `extra' observables, using multi-head self-attention, and generates the `context' vector. The second component is a conditional diffusion model, which, during training, learns to predict the noise added to the target parton-level distributions at each timestep, conditional on the context vector and a sinusoidal time-embedding, and inverts the noising process to reconstruct the parton-level distributions (and hence, the masses of $\lsptwo$ and $\lspone$) starting from a gaussian noise during inference. To enable a single network to perform mass reconstructions over a wide range of electroweakino masses, we train this combined framework on signal events sampled from a grid of benchmark points within the currently allowed parameter space, which will also be accessible at the HL-LHC. 



In Section~\ref{sec:benchmark_model}, we discuss our benchmark scenario and region of parameter space explored in this study. The neural network architecture and how it processes the event information are described in Sec.~\ref{sec:ML_model}. We discuss our results in Sec.~\ref{sec:result}, followed by an outlook and summary in Sec.~\ref{sec:summary}.

\section{The UDD Benchmark scenario}
\label{sec:benchmark_model}

In a previous study~\cite{Barman:2020azo}, the projected sensitivity of the signal process in Eqn.~\eqref{eq:eqn:signal_proc} at the HL-LHC has been explored. The analysis in \cite{Barman:2020azo} considered a final state with one isolated lepton ($\ell\equiv e,\mu$), exactly two photons, and at least two light-flavored jets, $pp \to \chonepm \lsptwo \to 1\ell + 2\gamma + jets$. The projected $2\sigma$ contours reach roughly up to $m_{\lsptwo/\chonepm} \sim 700~$GeV for $m_{\lspone} \sim 50~$GeV, falling down to $m_{\lsptwo/\chonepm} \sim 650~$GeV at $m_{\lspone} \sim 350~$GeV. Through a luminosity scaling to the current LHC, with $\mathcal{L} \sim 140~\mathrm{fb}^{-1}$, the $2\sigma$ contour excludes $m_{\lsptwo/\chonepm} \lesssim 350~$GeV for $m_{\lspone} \sim 50~$GeV. Hence, a wide region of parameter space with wino-like $\lsptwo/\chonepm$ masses ranging between $350~\mathrm{GeV} \lesssim m_{\lsptwo/\chonepm} \lesssim 650~\mathrm{GeV}$ will become available through searches in the $pp \to \chonepm \lsptwo \to 1\ell + 2\gamma + jets$ channel at the HL-LHC, irrespective of bino-like $\lspone$ masses.

For training, we generate event samples for $m_{\lsptwo/\chonepm}$ between 400~GeV to 650~GeV, with a step-size of $30~$GeV, and $m_{\lspone}$ varying between 50~GeV to $m_{\lsptwo/\chonepm} - m_h$, with a step-size of 25~GeV. We use \texttt{Pythia6} \cite{Sjostrand:2006za} to generate events at the hard-scattering level, followed by a simulation of showering and hadronization effects. \texttt{Delphes-3.5.0} \cite{deFavereau:2013fsa} is used for fast detector simulation, with the default ATLAS configuration card. We follow the event selection criteria from Ref.~\cite{Barman:2020azo}.




\section{Diffusion-based Reconstruction Framework}
\label{sec:ML_model}

We combine two different components in our network, a detector encoder and a diffusion model, that work in pairs. We first describe the architecture of the detector encoder, whose main role is to map the detector–level features into a permutation-invariant context vector $\mathcal{C}$, picked up later by the diffusion model as a conditional input. 


The following detector-level observables are considered:
\begin{eqnarray}
        &\{p_x, p_y, p_z, E\}_i \,:\, i = \ell, \gamma_1, \gamma_2,\, \& \, j_n~(2 \leq n \leq 4)\,& \nonumber \\
        &\cancel{E}_T, \Delta R_{min} (\ell, j_n),\, \Delta R_{max} (\ell, j_n),\, \Delta R_{min} (\ell, \gamma),\, \Delta R_{max} (\ell, \gamma),& \nonumber \\
        &\Delta R_{min} (\gamma,j_n),\, \Delta R_{max} (\gamma, j_n),\, \sum_{i=1}^n p_T(j_i).&
        \label{eq:det_obs}
\end{eqnarray}

We encode the four-momentum of the lepton, the two photons, and up to four jets into separate tokens, one for each final state particle. We also consider an extra token containing the additional variables in Eq.~\eqref{eq:det_obs}. Each of the $N_{tok} = 8$ tokens is mapped into a $d_{L}$ dimensional latent space using a linear layer followed by Rectified Linear Unit~(ReLU) activation. We pass the stacked sequence of latent representations, which has a shape of $N_{tok} \times d_{L}$, to the Transformer Encoder with 4 identical layers~\cite{vaswani2017attention}. Each layer performs multi-head self-attention~(number of heads: 4), followed by a feed-forward network with two hidden layers of inner dimension $d_{ff}$. Layer normalization and residual connections are employed in both the attention and feed-forward blocks, as described in Ref.~\cite{vaswani2017attention}. We also impose a dropout rate of 0.1 in both layers for training regularization. The output from the Transformer Encoder is $N_{tok} \times d_L$ shaped, which is then compressed into a $d_L$ dimensional context vector $\mathcal{C}$ by averaging over the $N_{tok}$ tokens. We note that $\mathcal{C}$ encodes the detector-level in a permutation-invariant way and will be used by the diffusion model as a conditional input. 

During the forward-noising process, we choose a small variance $\beta_i$ that linearly increases from $10^{-5}$ to 0.02 over $T = 1500$ time steps. This variance controls the amount of noise added to the parton-level vectors $x_p^0$ at each timestep $t$, where $t$ ranges from $0$ to $T$. The `noised' vectors at time step $t$ are given by, 
\begin{equation}
    x_p^t = \sqrt{\bar{\alpha}_t} \, x_p^0 + \sqrt{1 - \bar{\alpha}_t} \, \epsilon, 
    \label{eqn:noisy_target}
\end{equation}
where, $\bar{\alpha}_t$ is the cumulative product of signal retention factors $\alpha_i = 1 - \beta_i$, until time step $t$, 
\begin{equation}
    \bar{\alpha}_t = \prod_{i=1}^{t} \alpha_i,
\end{equation}
and $\epsilon$ is a random noise sampled from a Gaussian distribution $\mathcal{N}$ with a mean and variance of 0 and 1, respectively. During the initial steps $t$, $x_p^t$ still resembles the target parton-level vectors. For example, at $t = 50$, $\bar{\alpha}_t$ is around 0.987. But as $t$ increases, $\bar{\alpha}_t$ starts falling sharply. As $t$ approaches $T=1500$, $\bar{\alpha}_t \sim 10^{-5}$, and $x^T_p$ is almost identical to the pure noise $\epsilon$. Thus, the amount of noise heavily depends on the timestep, and during inference, the reverse diffusion process must be conditional on the timestep. Therefore, we map every integer value $t$ into a $d_t$-dimensional sinusoidal position embedding~\cite{vaswani2017attention}, 
\begin{equation}
    (pe)_{d_t} = \left[\sin\left(t\, \omega_0\right),\,\cos\left(t\, \omega_0\right),\,\sin\left(t\, \omega_1\right),\,\cos\left(t\, \omega_1\right),..., \sin\left(t\, \omega_{\frac{d_t-1}{2}}\right),\,\cos\left(t\, \omega_{\frac{d_t-1}{2}}\right)  \right],
\end{equation}
where the frequency $\omega_k$ is chosen from~\cite{vaswani2017attention}, 
\begin{equation}
    \omega_k = \left(\frac{1}{10000}\right)^{\frac{k}{d_t/2 - 1}}
\end{equation}

\begin{figure}[!t]
    \centering
    \includegraphics[width=\linewidth]{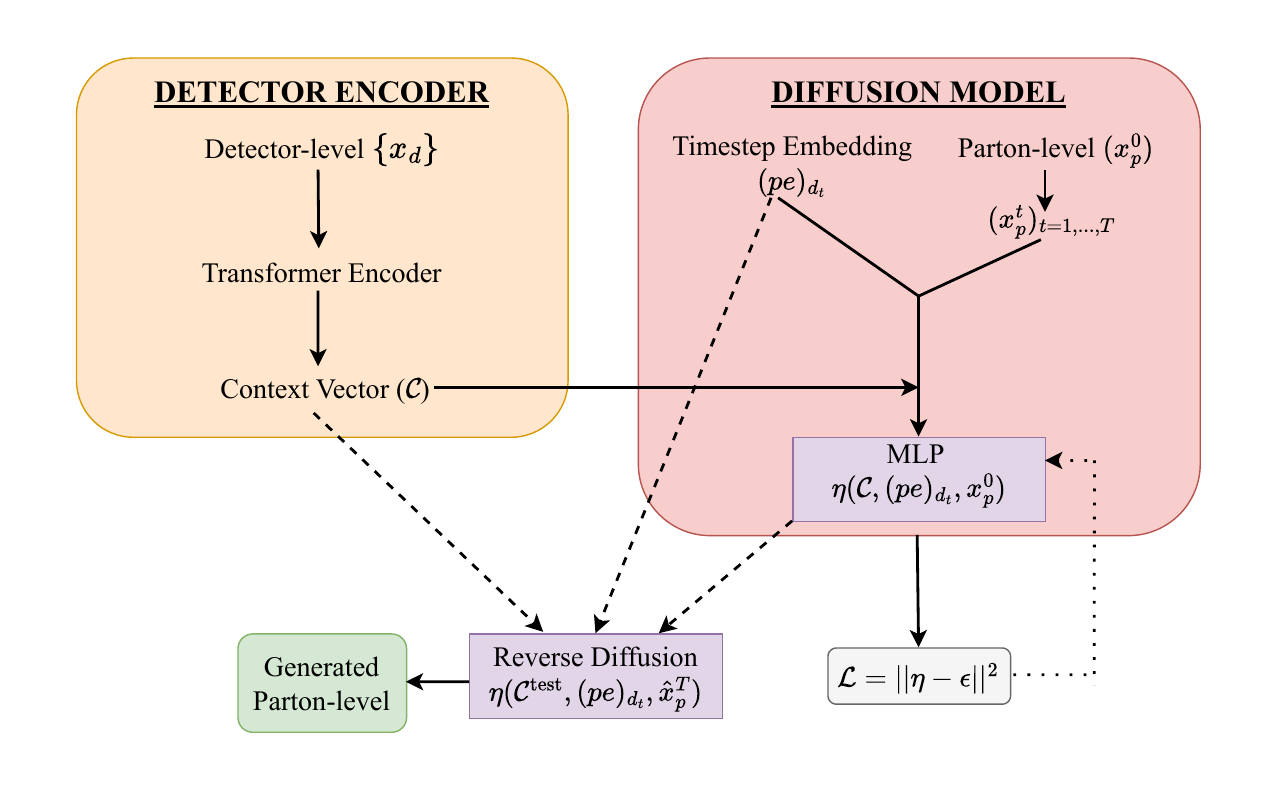}
    \caption{A schematic representation of the model architecture is shown.}
    \label{fig:model_architecture}
\end{figure}

During training, the time-embedding $(pe)_{d_t}$ and the context vector $\mathcal{C}$ from the detector encoder are both independently transformed into a $d_\mathrm{DF}$ dimensional latent vector space using a two-layer fully connected NN with ReLU activation. They are concatenated with the `noised' parton-level vectors $x_p^t$ at a timestep $t$ generated using Eqn.~\eqref{eqn:noisy_target}, which are then passed through an MLP with three hidden layers, each followed by Gaussian Error Linear Unit~(GELU) activation~\cite{hendrycks2023gaussianerrorlinearunits}. For every event in the minibatch, the noisy target $x_p^t$ is generated by choosing a random value of time step $t$, and a noise vector $\epsilon$. The MLP maps the concatenated $2d_\mathrm{DF} + 8$ dimensional input to an $8$ dimensional output $\eta$, which is the network's prediction for the noise. We adopt a mean-squared error loss between the predicted noise $\eta$, that depends on $\mathcal{C}$, $(pe)_{d_t}$ and $x_p^t$, and the true noise $\epsilon$: $\mathcal{L} = \mathbb{E}_{x_p^0, t, \epsilon} || \epsilon - \eta(\mathcal{C}, (pe)_{d_t},x_p^t)||^2$. The network architecture is illustrated in Fig.~\ref{fig:model_architecture}.

The network learns the amount of noise injected at each timestep $t$, conditioned on $(pe)_{d_t}$ and the detector-level features. During testing, the reverse-diffusion process begins at timestep $T$ with $\hat{x}_p^T$ chosen from the Gaussian noise $\hat{x}_p^T \sim \mathcal{N}(0, 1)$. The previous state $x_p^{T-1}$ is computed as~\cite{ho2020denoising}, 
\begin{equation}
    \hat{x}_p^{T-1} = \mu_T + \sqrt{\beta_T}\, z,
\end{equation}
where $\mu_T$ is the posterior mean at step $T$, 
\begin{equation}
    \mu_T = \frac{1}{\sqrt{\alpha_T}}\left(\hat{x}_p^{T} - \frac{1 - \alpha_T}{\sqrt{1 - \bar{\alpha}_T}}\,  \eta\left( \mathcal{C}^{\mathrm{test}}, (pe)_{d_t},\hat{x}_p^T\right)\right), 
    \label{eqn:posterier_mean}
\end{equation}
and $z$ is a small Gaussian noise. In Eqn.~\eqref{eqn:posterier_mean}, $\mathcal{C}^{\mathrm{test}}$ is the context vector corresponding to the detector-level events in the test dataset $x_d^{\mathrm{test}}$, and $\eta (\mathcal{C^{\mathrm{test}}}, (pe)_{d_t},\hat{x}_p^T)$ is the noise predicted by the network at time step $T$. This process is repeated for time-steps $t = T-1,\,T-2\,..., 2$. At the final time step, we set $\hat{x}_p^0 = \mu_1$, where $\hat{x}_p^0$ are the model-generated parton-level distributions. 
\begin{table}[!t]
    \centering
    \begin{tabular}{|l|c|} \hline
        \multicolumn{1}{|c|}{Hyperparameters} & Value  \\ \hline \hline 
        Dimensionality of latent representation for tokens: $d_L$ & 256 \\
        Number of Encoder layers & 4 \\
        Number of heads & 4 \\
        Inner dimension of feed-forward networks: $d_{ff}$ & 256 \\
        Dropout & 0.1 \\ \hline 
        Time steps in forward noising process: $T$ & 1500 \\
        Dimensionality of context vector and timestep embeddings: $d_{\mathrm{DF}}$ & 256 \\ 
        Output dimensions & 8 \\ \hline 
        Optimizer & Adam \\
        Learning Rate & $10^{-4}$ \\ \hline 
    \end{tabular}
    \caption{Hyperparameters for our diffusion-based reconstruction framework.}
    \label{tab:hyperparamter}
\end{table}

We implement our model in the PyTorch framework~\cite{paszke2019pytorchimperativestylehighperformance}. The network is trained with the Adam optimizer, and we choose a learning rate of $10^{-4}$. We summarize the network hyperparameters in Table~\ref{tab:hyperparamter}.  

\section{Results}
\label{sec:result}
We first study the performance of our network on a benchmark signal point with $m_{\lsptwo}=600~$GeV and $m_{\lspone}=200~$GeV. 
\begin{figure}[!htb]
    \centering
    \includegraphics[width=0.45\linewidth]{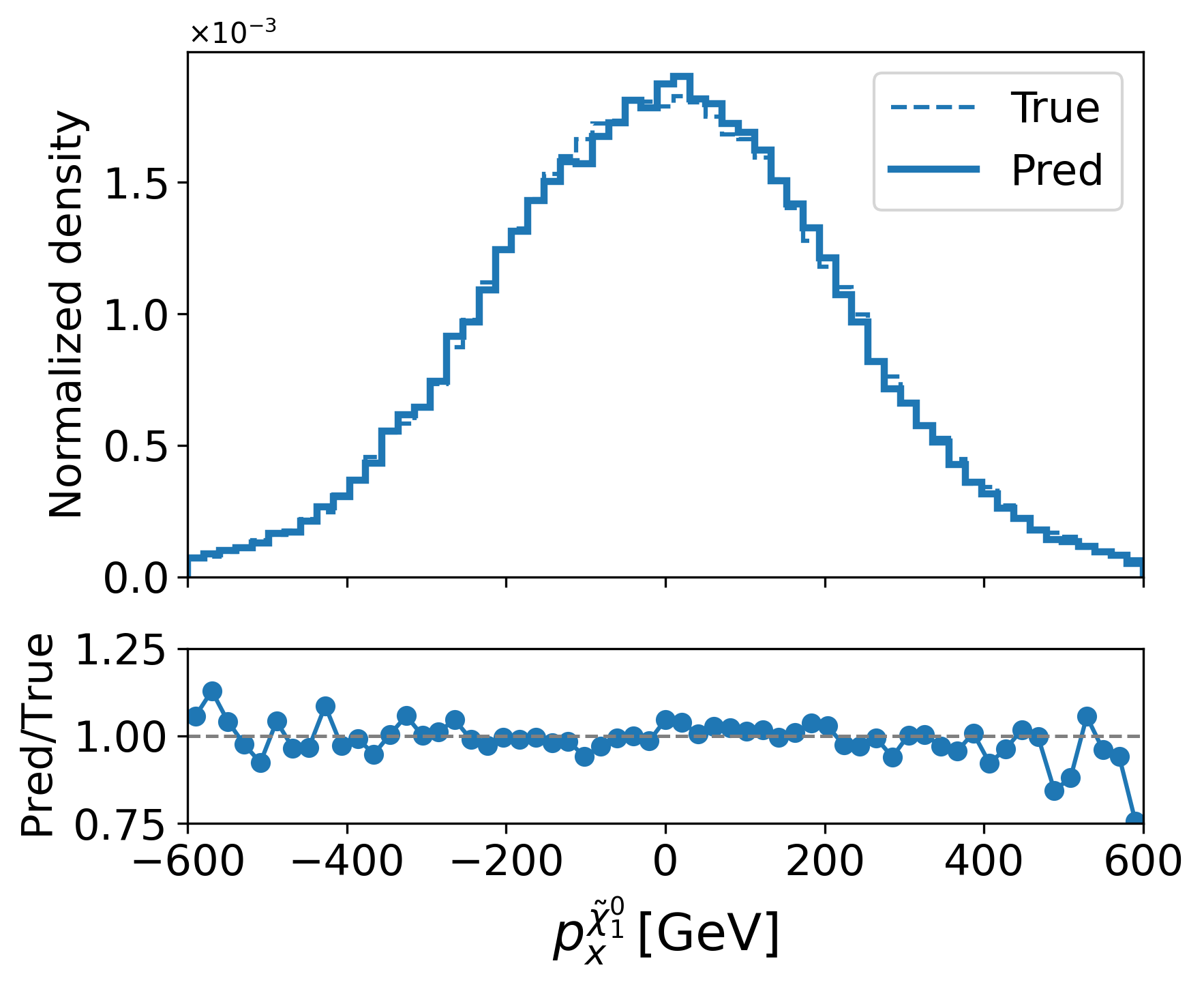}\hspace{0.5cm}\includegraphics[width=0.45\textwidth]{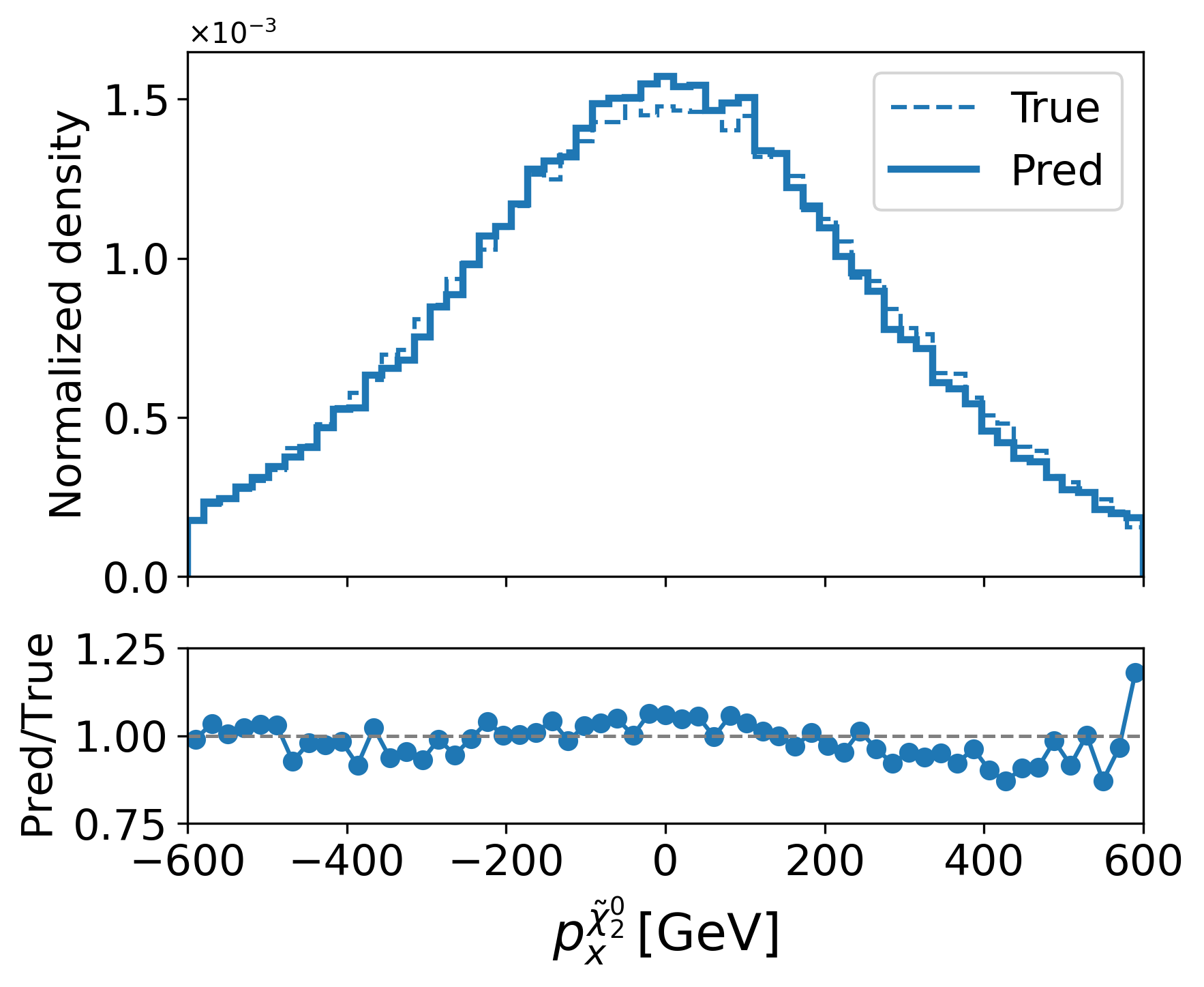}\\
    \includegraphics[width=0.45\linewidth]{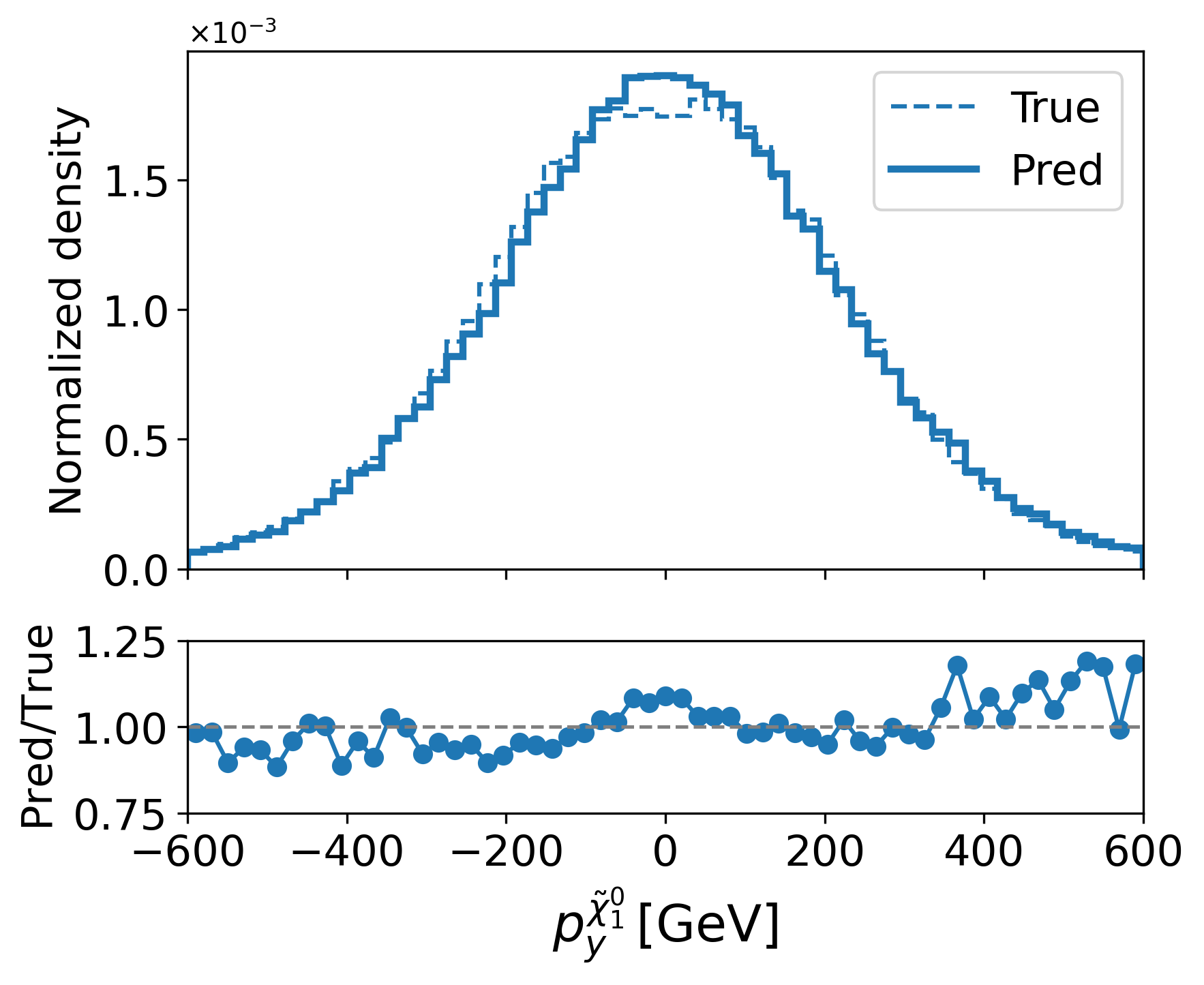}\hspace{0.5cm}\includegraphics[width=0.45\textwidth]{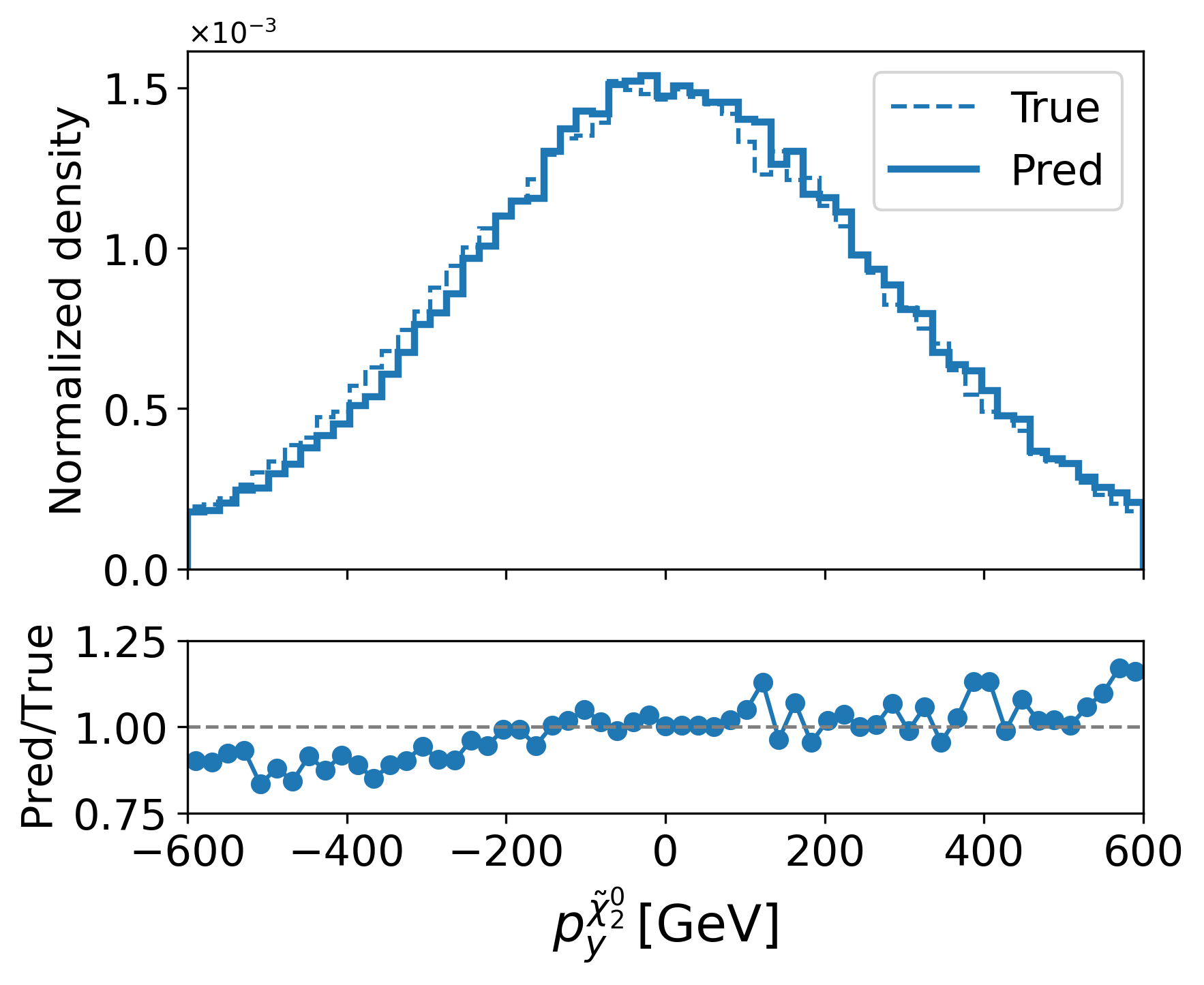}\\
    \includegraphics[width=0.45\linewidth]{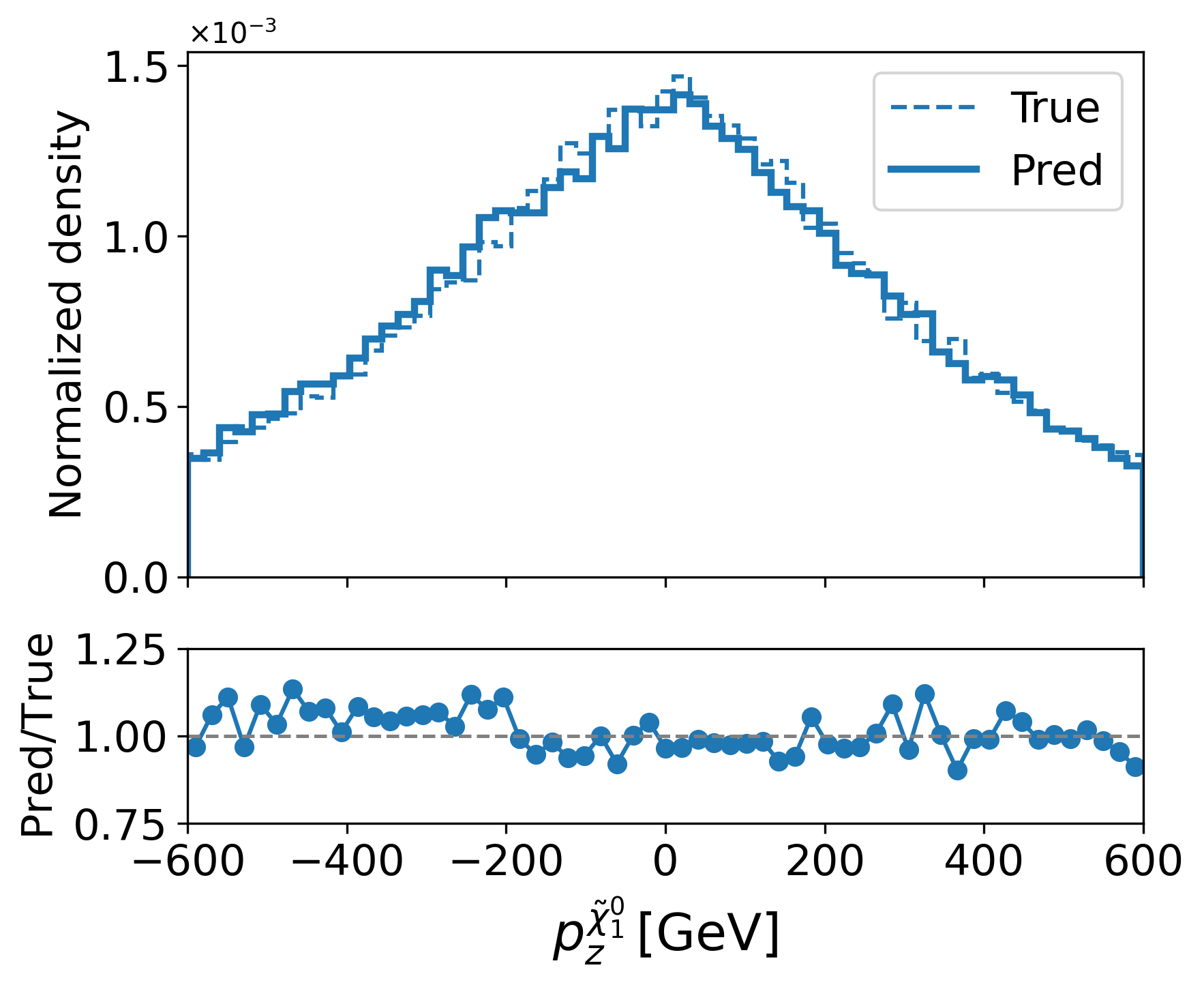}\hspace{0.5cm}\includegraphics[width=0.45\textwidth]{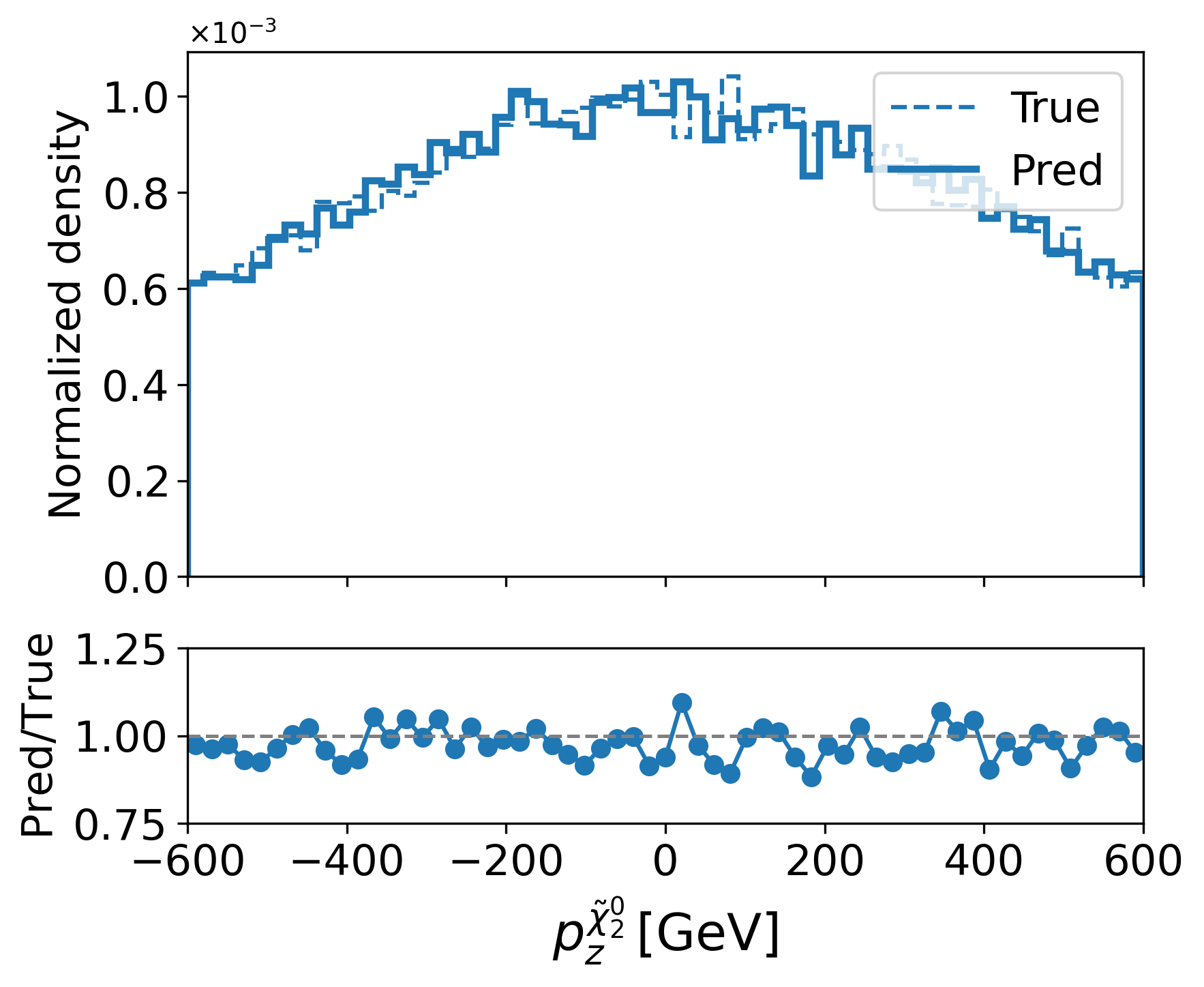}
    \caption{The generated~(`Pred') and truth~(`True') parton-level distributions for $p_{x},\, p_y,\,p_z$ for the lightest neutralino $\lspone$~(left) and the next-to-lightest-neutralino $\lsptwo$~(right) are shown for the process $pp \to \chonepm\lsptwo \to (\chonepm \to W^{\pm} \lspone) (\lsptwo \to h \lspone)$ at the $\sqrt{s}=14~$TeV LHC, with $m_{\lsptwo} = 600~$GeV and $m_{\lspone}=200~$GeV.}
    \label{fig:fixed_point}
\end{figure}
The network is trained on $1.2 \times 10^5$ event samples, for 100 epochs using a batch size of 128, and is applied on a test dataset for the same signal benchmark. 
We show the true and generated parton-level distributions for the $p_x$, $p_y$, and $p_z$ of the next-to-lightest neutralino $\lsptwo$ and its daughter $\lspone$ in Fig.~\ref{fig:fixed_point}. We observe that the distributions generated by the network have a good overlap with the true parton-level across the bulk of phase space. This is indicated by the respective lower panels, where we show the ratio of the generated parton-level to the true parton-level, which remains roughly within $10-15\%$ except in the tails of the distribution, where larger deviations are observed due to lower event statistics.

\begin{figure}[!htb]
    \centering
    \includegraphics[width=0.6\linewidth]{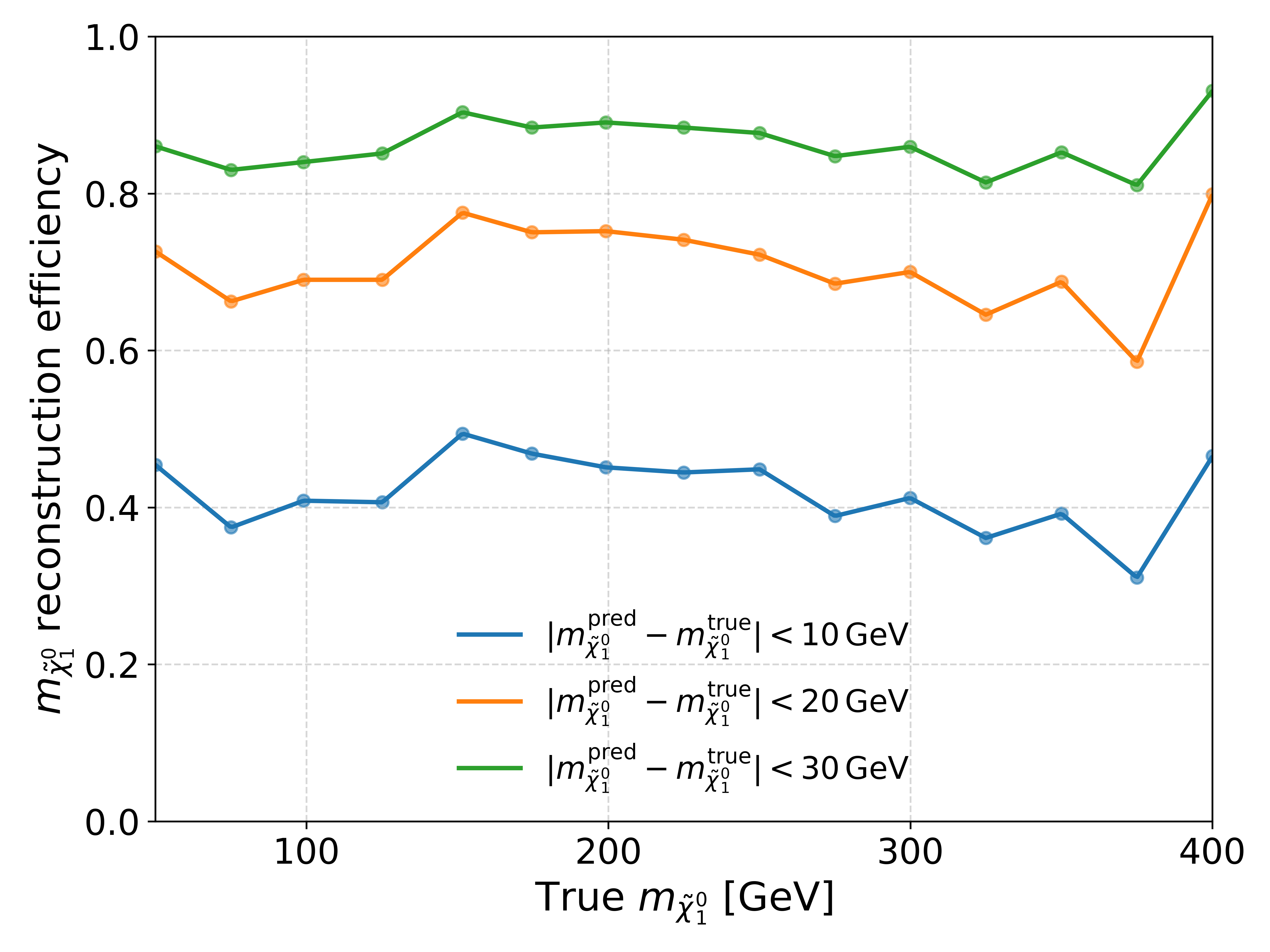}
    \caption{The reconstruction efficiencies, defined as the fraction of events where $|m_{\lspone}^{\mathrm{pred}} - m_{\lspone}^{\mathrm{true}}| < \Delta m$, are shown, for three tolerances, $\Delta m = 10,\,20$ and 30~GeV. Here, the network is trained on events with a fixed $m_{\lsptwo} = 600~$GeV, but with $m_{\lspone}$ varying between 50 and 400~GeV, with a 50~GeV step size. The network is evaluated on events generated with $m_{\lspone}$ varying over the same range, but with a smaller step-size of 25~GeV.}
    \label{fig:reco_eff_nlsp_600GeV}
\end{figure}

Next, we investigate the network's ability to interpolate across different values of $m_{\lspone}$, keeping the mass of $\lsptwo$ fixed at $m_{\lsptwo} = 600~$GeV. To this end, we train the network on signal events generated with $m_{\lspone}$ ranging between 50~GeV to 400~GeV, with a 50~GeV step size. We consider 35000 events from each benchmark point. Here again, the network is trained for 100 epochs with a batch size of 128. To test the network's performance, we consider signal events generated with $m_{\lspone}$ varying between the same range, 50~GeV to 400~GeV, but with a step-size of 25~GeV. For each signal benchmark in the test dataset, we evaluate the reconstruction efficiency of $m_{\lspone}$, defined as the fraction of events where the difference between the generated mass of the lightest neutralino $m_{\lspone}^{\textrm{pred}}$ and its true mass, represented here as $m_{\lspone}^{\textrm{true}}$, lies within a tolerance $\Delta m$, $ |m_{\lspone}^{\mathrm{pred}} - m_{\lspone}^{\mathrm{true}}| < \Delta m$. In Fig.~\ref{fig:reco_eff_nlsp_600GeV}, we present the reconstruction efficiencies for three different tolerances, $\Delta m = 10,\, 20$, and 30~GeV. 
For $\Delta m = 30~$GeV, our network achieves a reconstruction efficiency of $\gtrsim 80\%$ over the entire range $50~\mathrm{GeV} \leq m_{\lspone} \leq 400~\mathrm{GeV}$, even for intermediate mass points not included in the training dataset. The reconstruction efficiency drops to $\gtrsim 60\%$ over the entire $m_{\lspone}$ range, for a smaller tolerance $\Delta m = 20~$GeV. These results indicate that the network successfully interpolates across the wide parameter space and generalizes well to reconstruct previously unseen mass points.

\begin{figure}[!htb]
    \centering
    \includegraphics[width=0.47\linewidth]{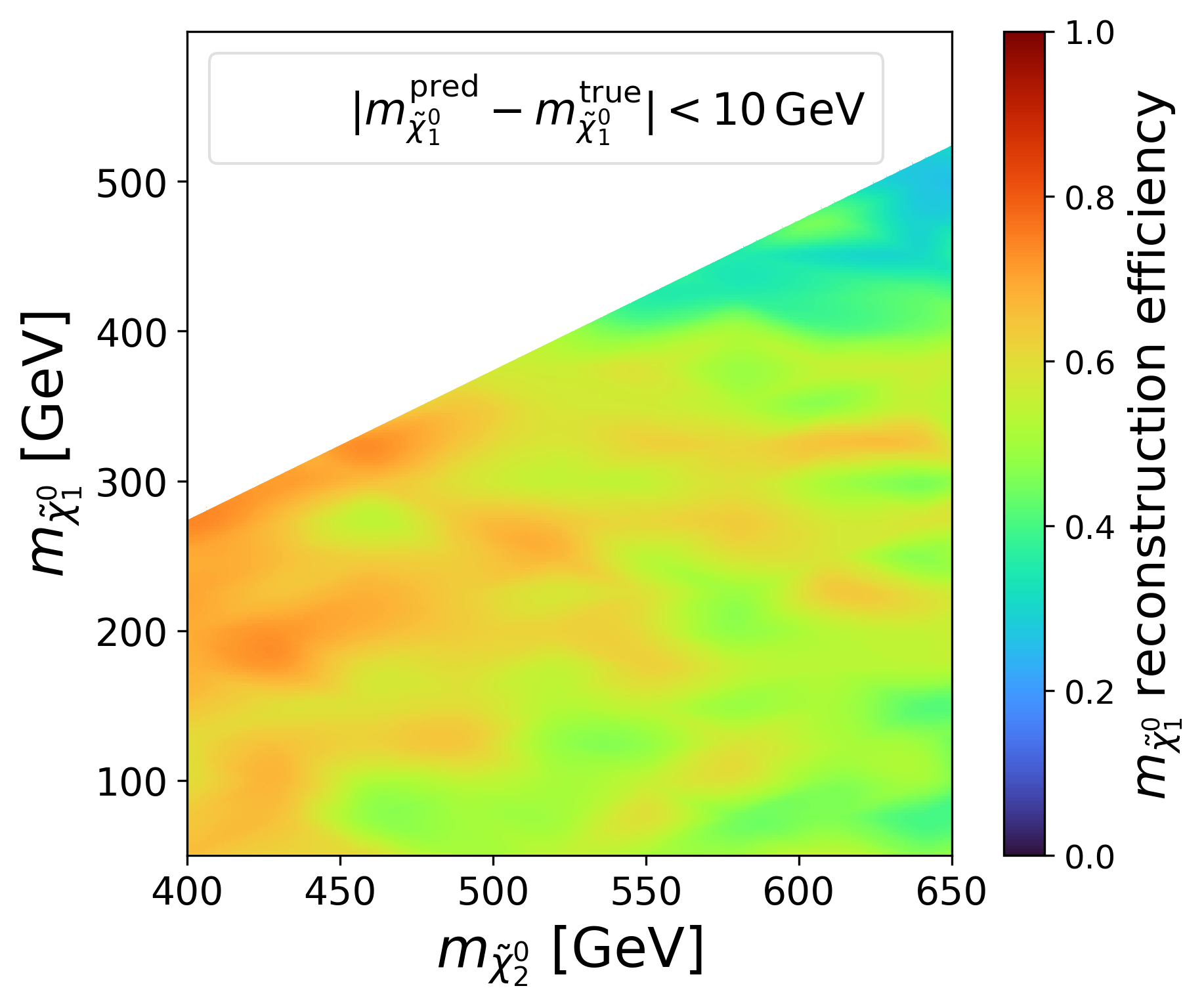}\hspace{0.3cm}\includegraphics[width=0.47\linewidth]{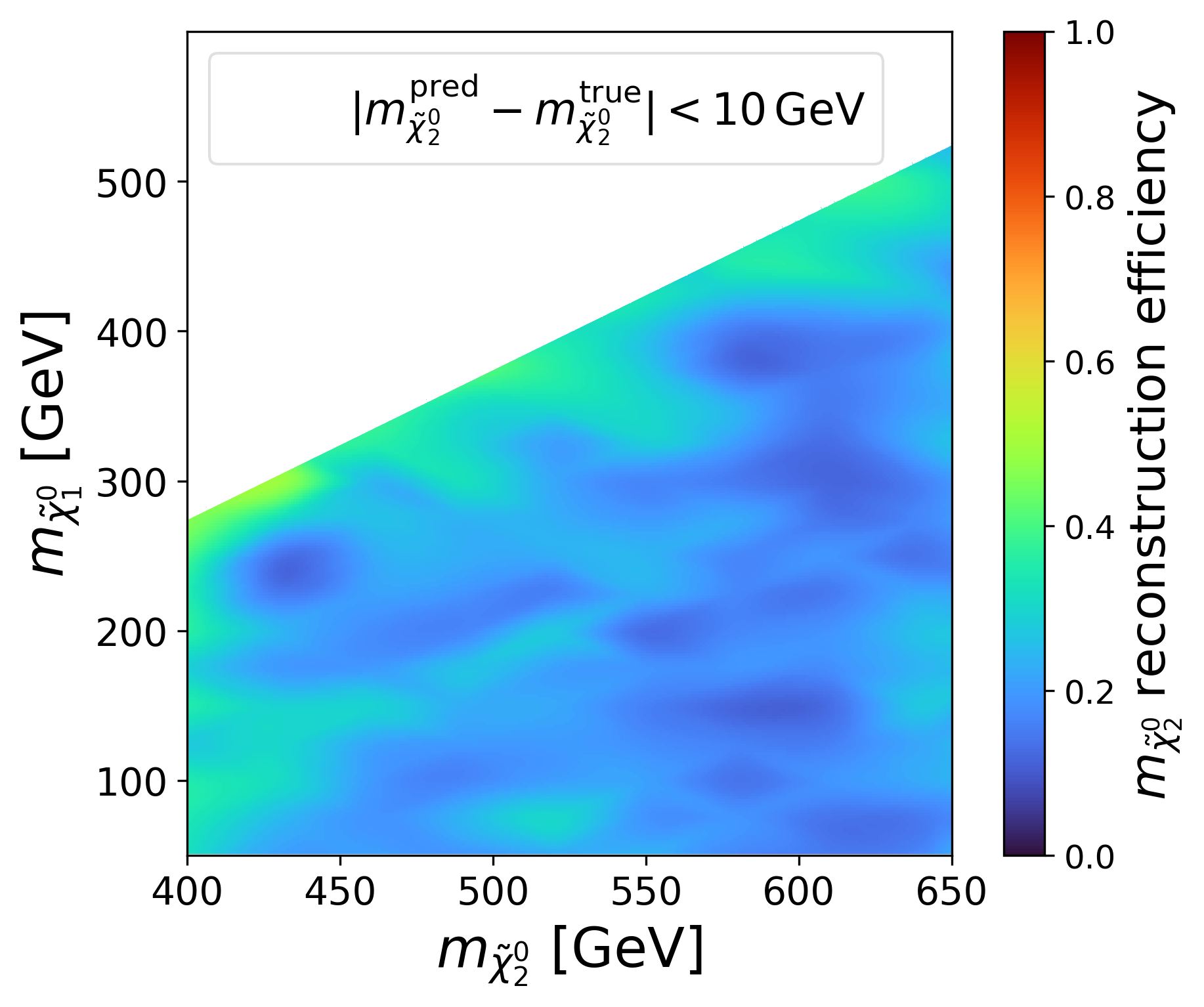}\\
    \includegraphics[width=0.47\linewidth]{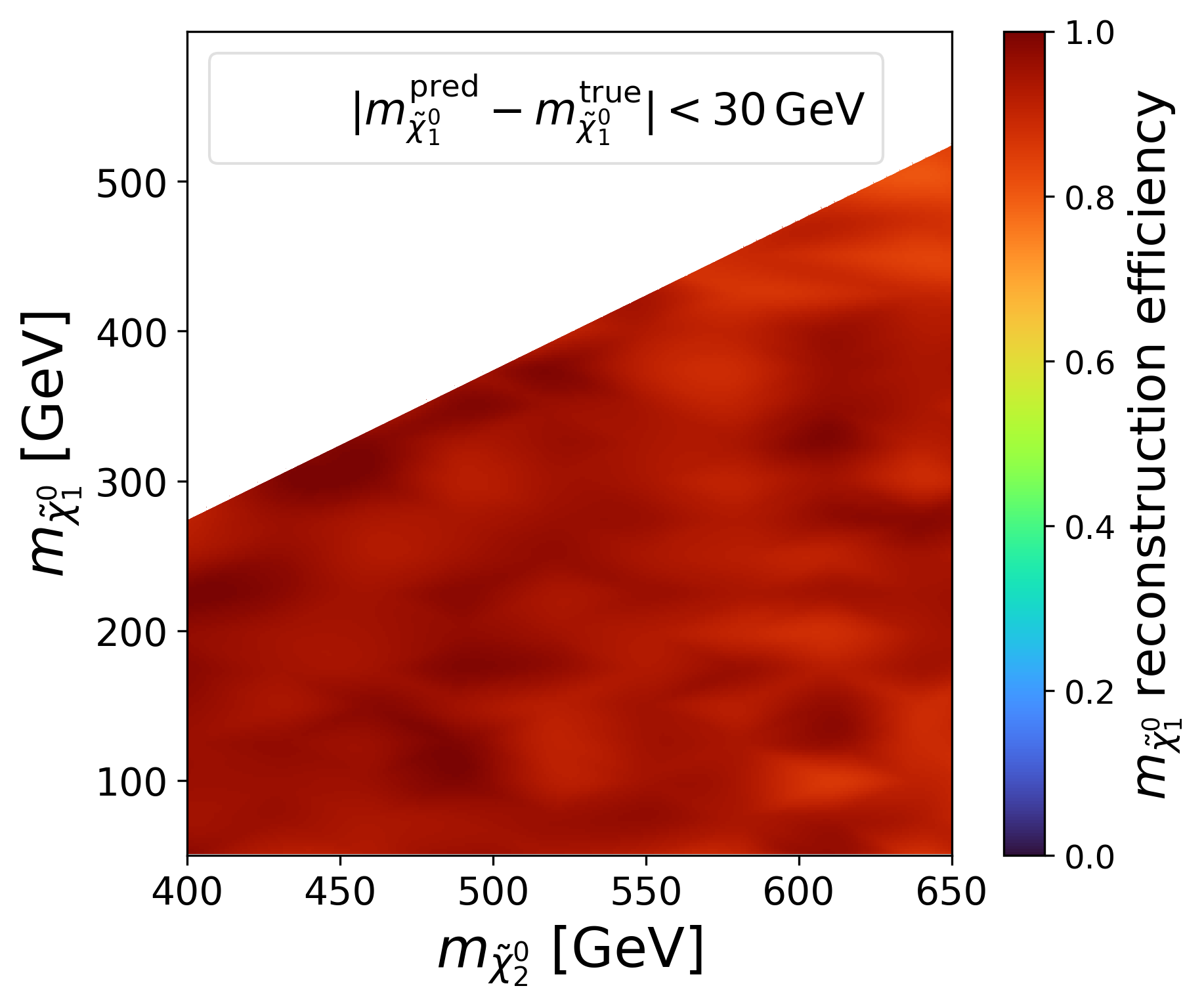}\hspace{0.3cm}\includegraphics[width=0.47\linewidth]{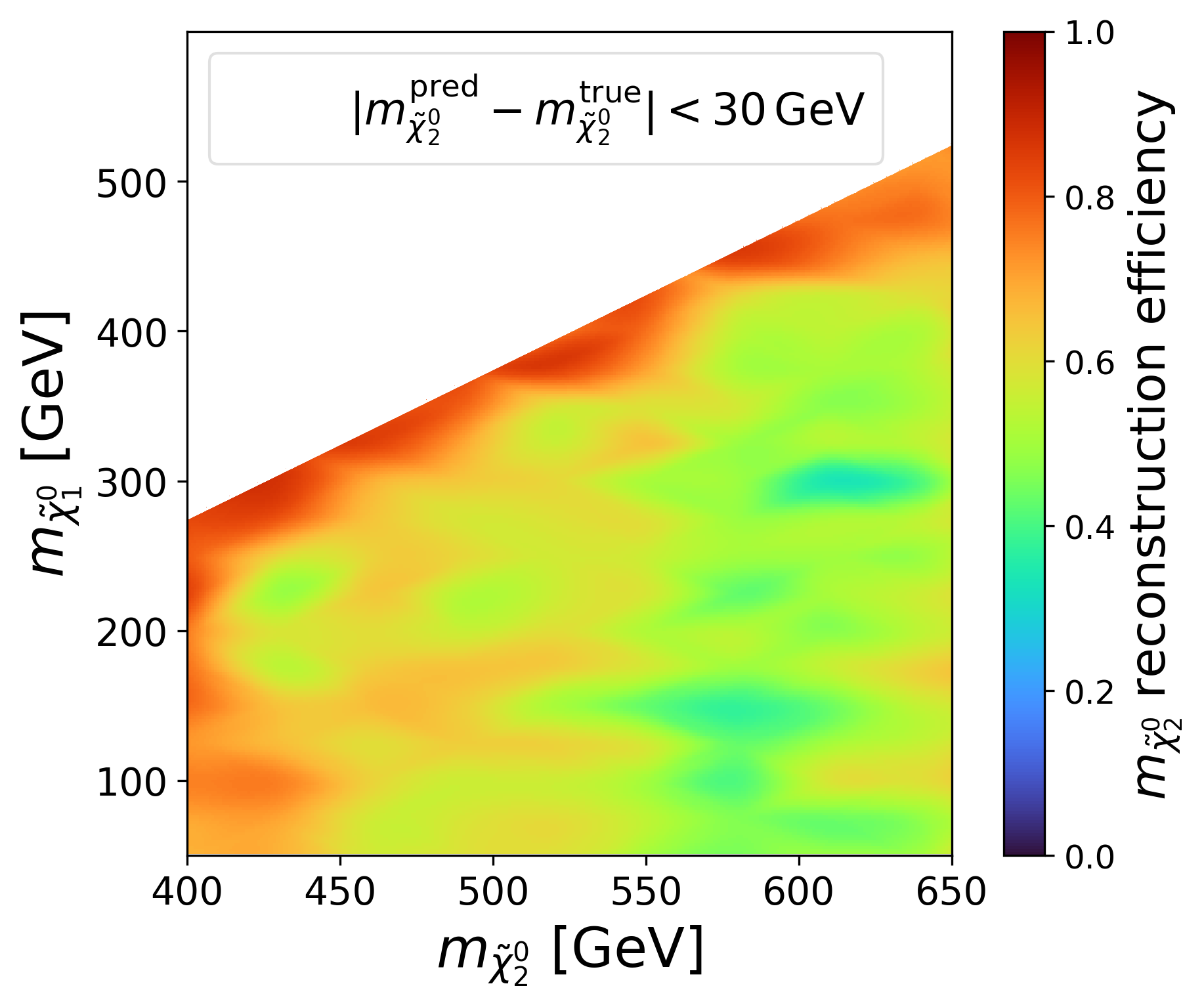}\\
    \includegraphics[width=0.47\linewidth]{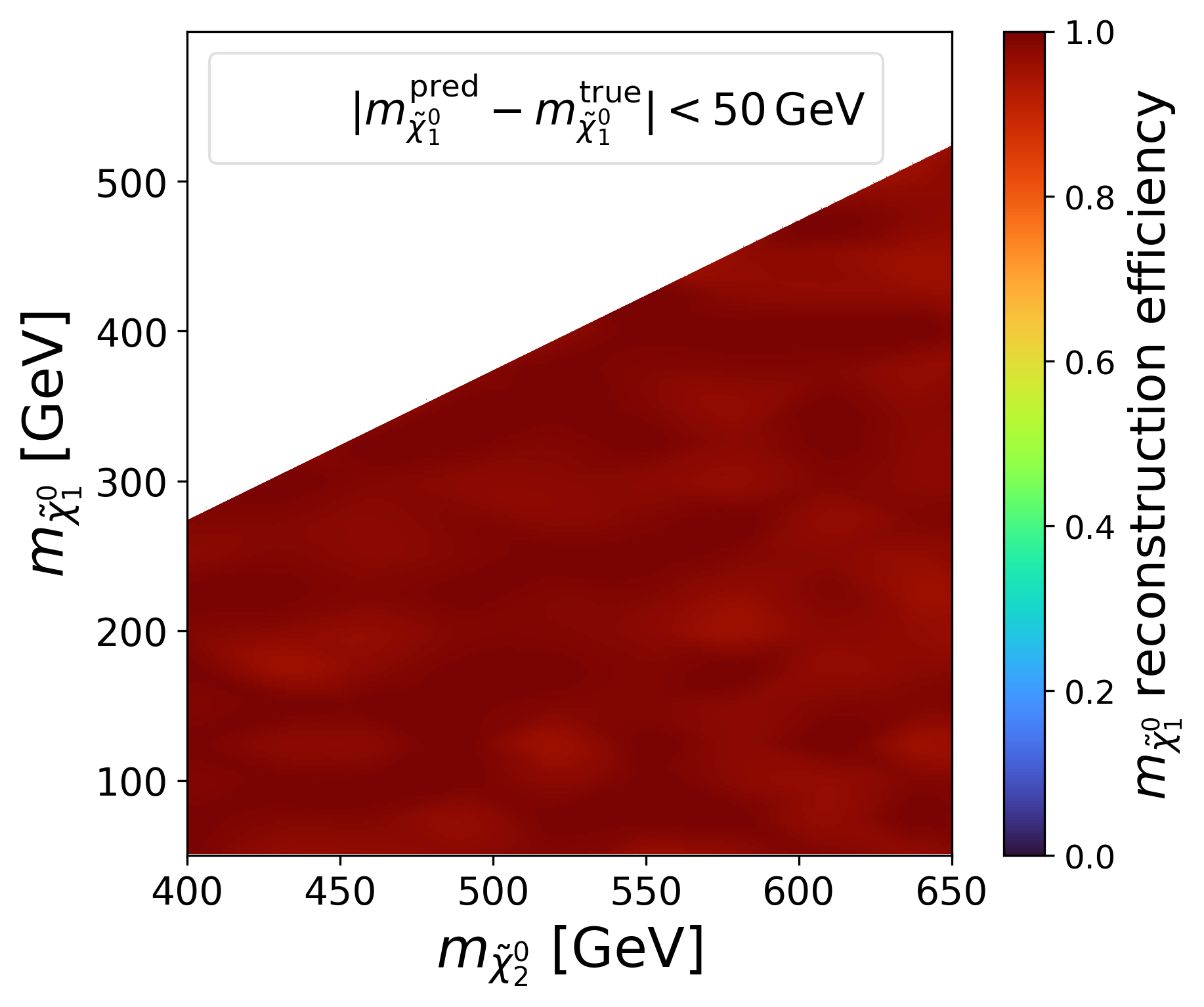}\hspace{0.3cm}\includegraphics[width=0.47\linewidth]{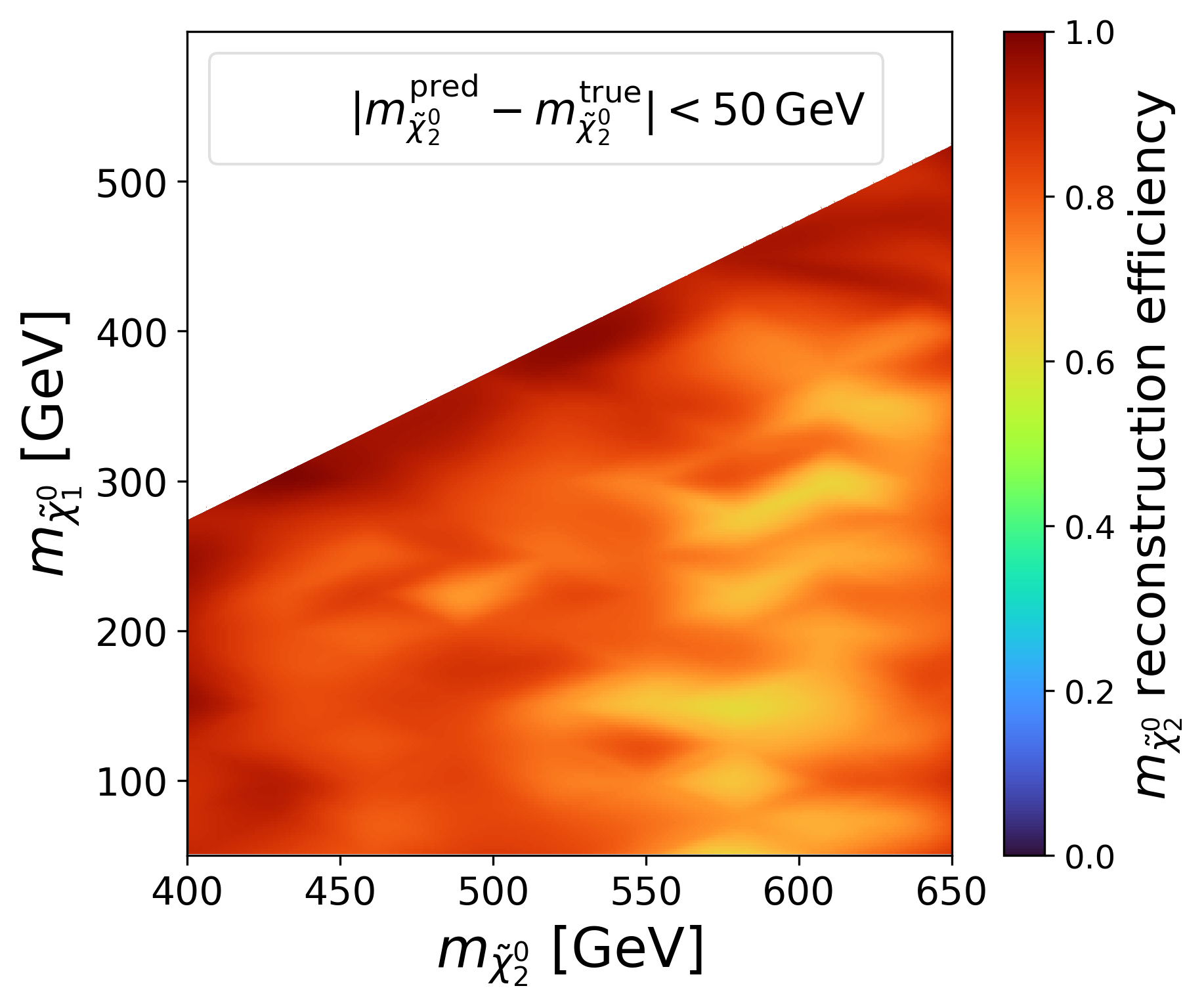}\\
    \caption{The reconstruction efficiencies for $m_{\lspone}$~(left) and $m_{\lsptwo}$~(right) are shown for three different tolerances, $\Delta m = 10,\,30$ and 50~GeV. The network is trained on events with $m_{\lsptwo}$ ranging between 400 to 650~GeV with a 30~GeV step-size, and $m_{\lspone}$ varying between 50~GeV to $m_{\lsptwo} - 125~$GeV with a 25~GeV step-size. The network is evaluated on different events sampled from the same mass grid.}
    \label{fig:heatmap}
\end{figure}

Finally, we examine how well the network generalizes over the full parameter space spanning both $m_{\lsptwo}$ and $m_{\lspone}$. We train the network on signal events drawn from a $\{m_{\lsptwo},m_{\lspone}\}$ grid: $m_{\lsptwo}$ ranges from 400~GeV to 650~GeV, with step-size 30~GeV, while $m_{\lsptwo}$ ranges between 50~GeV to $m_{\lsptwo} - m_h$ with step-size 25~GeV. We consider 5500 events from each mass point. The network is trained over 100 epochs with a batch size of 256, and tested on event samples generated over the training mass grid. In Fig.~\ref{fig:heatmap}, we show the reconstruction efficiencies for $m_{\lspone}$~(left panel) and $m_{\lsptwo}$~(right panel), for three different tolerances, $\Delta m = 10,\,30$ and 50~GeV. At the tightest tolerance, $\Delta m = 10~$GeV, the network performs rather weakly with the $m_{\lsptwo}$ reconstruction efficiency barely reaching around $40\%$ only in some regions of the parameter space. The reconstruction efficiency for $m_{\lspone}$ turns out to be comparatively better, with values reaching roughly up to $70\%$ in the smaller $m_{\lsptwo}$ region. As we relax the tolerances, the performance improves noticeably. At $\Delta m = 30~$GeV, the $m_{\lspone}$ reconstruction efficiency stays roughly above $\gtrsim 70\%$, while the efficiency for $m_{\lspone}$ remains above $\gtrsim 40\%$, over the entire parameter space. Finally, at $\Delta m = 50~$GeV, the network achieves roughly $\gtrsim 90\%$ efficiency for $m_{\lspone}$, while the efficiency for $m_{\lsptwo}$ stays above $60\%$ over the bulk of phase space.

These results demonstrate the network's ability to generalize across the masses of both the lightest $\lspone$ and next-to-lightest neutralino $\lsptwo$, even though the reconstruction efficiency for $m_{\lsptwo}$ in particular could benefit from further improvements. Given the broad parameter space considered here and relying on inclusive detector-level inputs, a reconstruction efficiency of $40-50\%$ for $m_{\lsptwo}$ and an efficiency of $\gtrsim 70\%$ for $\lspone$, at intermediate $\Delta m = 30~$GeV, is a promising step towards constructing more robust and model-agnostic mass inference frameworks. We observe that the reconstruction efficiency for $m_{\lsptwo}$ is lower than $m_{\lspone}$. This is expected since the $\lspone$ decays promptly into three jets, while the $\lsptwo$ undergoes a longer cascade into the $h$ and the $\lspone$, making its mass reconstruction more challenging. We expect that adopting a denser grid of $\lsptwo$ masses in the training dataset, and including additional observables that are directly sensitive to the cascade decay topology of $\lsptwo$, could lead to improvements in the $m_{\lsptwo}$ reconstruction efficiencies. Additionally, adopting a hierarchical reconstruction approach, where the $m_{\lsptwo}$ reconstruction is conditioned on the predicted value of $m_{\lspone}$, might improve performance. Incorporating these enhancements is beyond the scope of the present study, and we leave their investigation to future work.

Before concluding this section, we would like to note that our diffusion-based reconstruction framework is not specific to this model and can be applied to other setups with different BSM scenarios and final-state topologies. As an example, we also test this method in an RPV-SUSY scenario with LLE-type $\lambda_{121}$ and/or $\lambda_{122}$ couplings, leading to a multi-lepton plus $\cancel{E}_T$ final state, and found comparable performance. In general, our framework can be applied to other search processes involving heavy resonant particles in the decay chain, to reconstruct their masses over a wide parameter region.

\section{Summary}
\label{sec:summary}

We introduced a two-component neural network, comprising a detector encoder that maps the detector-level inputs into a context vector via a Transformer, and a diffusion neural network that reconstructs the parton-level kinematics by reversing the learnt noising process conditioned upon the context vector. In this study, we investigate the potential of our diffusion-based ML framework to reconstruct the masses of $\lsptwo$ and $\lspone$ at the HL-LHC, presuming an excess in $pp \to \chonepm \lsptwo \to 1\ell + 2 \gamma + jets$ final state, within the R-parity violating MSSM scenario with $\lambda^{\prime\prime}_{112}$-type coupling. 

We first demonstrate the performance of the network at a single benchmark point ($\mlsptwo=600$ GeV, $\mlspone=200$ GeV). Here, we observe a good agreement between the generated and true parton-level distributions of $p_x$, $p_y$, and $p_z$ for both the lightest $\lspone$ and the second-lightest neutralino $\lsptwo$. Next, we vary $m_{\lspone}$ for a fixed $ m_{\lsptwo} = 600~$GeV and obtain $m_{\lspone}$ reconstruction efficiencies above $\gtrsim 80\%$ for a tolerance of $\Delta\, m =30$ GeV, over the entire $m_{\lspone}$ range, even at intermediate mass points not seen during the training. Finally, we train our neural network over the full two-dimensional mass grid, $400~\mathrm{GeV} \leq m_{\lsptwo} \leq 650~\mathrm{GeV}$ and $50~\mathrm{GeV} \leq m_{\lspone} \leq m_{\lsptwo} - m_h$. The network achieved a reconstruction efficiency of $\gtrsim 70\%$ and $\gtrsim 40\%$ for $m_{\lspone}$ and $\lsptwo$, respectively, over most of the parameter space, at a moderate $\Delta\, m = 30~$GeV. On further relaxing $\Delta\,m = 50~$GeV, the respective efficiencies improved to $\gtrsim 90\%$ and $\gtrsim 60\%$. To further illustrate the effectiveness and generalizability of our framework, we apply it to a different process, $pp \to \chonepm\chonemp + \chonepm\lsptwo \to 4\ell + \cancel{E}_T$ channel, which involves multiple sources of $\cancel{E}_T$ at the detector-level. In this case, the network achieves a $\mlspone$ reconstruction efficiency of $\gtrsim 80\%$ for $m_{\lsptwo}=1.5$~TeV and 2~TeV, for a tolerance of $\Delta m=30$ GeV. Overall, we note that our diffusion-based architecture can be easily adapted to various other BSM searches to uncover the underlying mass spectrum or parton-level kinematics, using detector-level observables.

\subsection*{Acknowledgement}
  The work of RKB is supported by the World Premier International Research Center Initiative (WPI), MEXT, Japan, and by JSPS KAKENHI Grant Number JP24K22876. Some of the computation for this work was performed using resources at Kavli IPMU. A. Choudhury acknowledges Anusandhan National Research Foundation (ANRF) India for the Core Research Grant no. CRG/2023/008570. S. Sarkar acknowledges Anusandhan National Research Foundation (ANRF) India for the financial support through the Core Research Grant No. CRG/2023/008570.

\vspace{0.5cm}
\noindent \textbf{Author contribution}:
All authors have contributed equally.

\vspace{0.5cm}
\noindent \textbf{Data availability statement:} No data associated in the manuscript.


\appendix
\section*{Appendix A: The LLE Benchmark scenario}
\begin{figure}[h]
    \centering
    \includegraphics[width=0.4\linewidth]{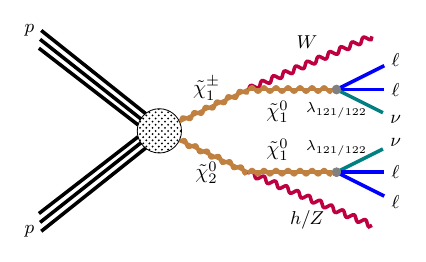} \hspace{1cm}
    \includegraphics[width=0.4\linewidth]{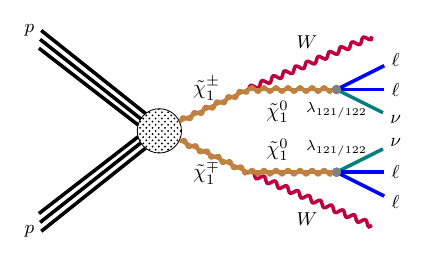}
    \caption{Diagrams for wino-like $\chonepm\lsptwo+\chonepm\chonemp$ pair-production and consequent decay of LSP via nonzero $\lambda_{121}$ and/or $\lambda_{122}$ coupling.}
    \label{fig:feyn_diag_LLE}
\end{figure}

For further testing of our ML model, we consider another benchmark scenario with lepton number violating nonzero $\lambda_{121}$ and/or $\lambda_{122}$ LLE type RPV coupling\footnote{For a complete list of $\lspone$ decay modes corresponding to various nonzero LLE couplings, please refer to Table.~II of \cite{Choudhury:2023eje}}. Here, directly pair-produced mass-degenerate wino-like chargino-chargino and chargino-neutralino ($pp\to \chonepm\chonemp +\chonepm\lsptwo$) goes to a cascade decay mode to bino-like LSP ($\lspone$) and $W$ and $Z/h$, followed by each $\lspone$ decaying to two leptons and one neutrino ($\lspone\to \ell\ell\nu,~\ell\equiv e,\mu$) via nonzero $\lambda_{121}$ and/or $\lambda_{122}$ coupling. 
With $\lambda_{121}\neq0$ and/or $\lambda_{122} \neq0$, the LSP pair decays to the $4\ell+\met$ configuration with $100\%$ branching ratio. Additional leptons may come from the decay of $W/Z/h$. Thus, the final state contains at least four leptons and $\met$, and we have at least two sources of $\met$ (See Fig.~\ref{fig:feyn_diag_LLE}). Hence, reconstructing the LSP mass here is more challenging than the UDD benchmark scenario.
The existing $2\sigma$ limit on the NLSP mass ($\mchonepm/\mlsptwo$) for this $N_{\ell}\geq4+\met$ at the LHC is $\sim1.6$ TeV for $\mlspone \sim 1$ TeV \cite{ATLAS:2021yyr}, and the projected $2\sigma$ reach at the HL-LHC is $\sim2.37$ TeV for $\mlspone\sim2.25$ TeV \cite{Choudhury:2023eje}.\\

\begin{figure}[!htb]
    \centering
    \includegraphics[width=0.47\linewidth]{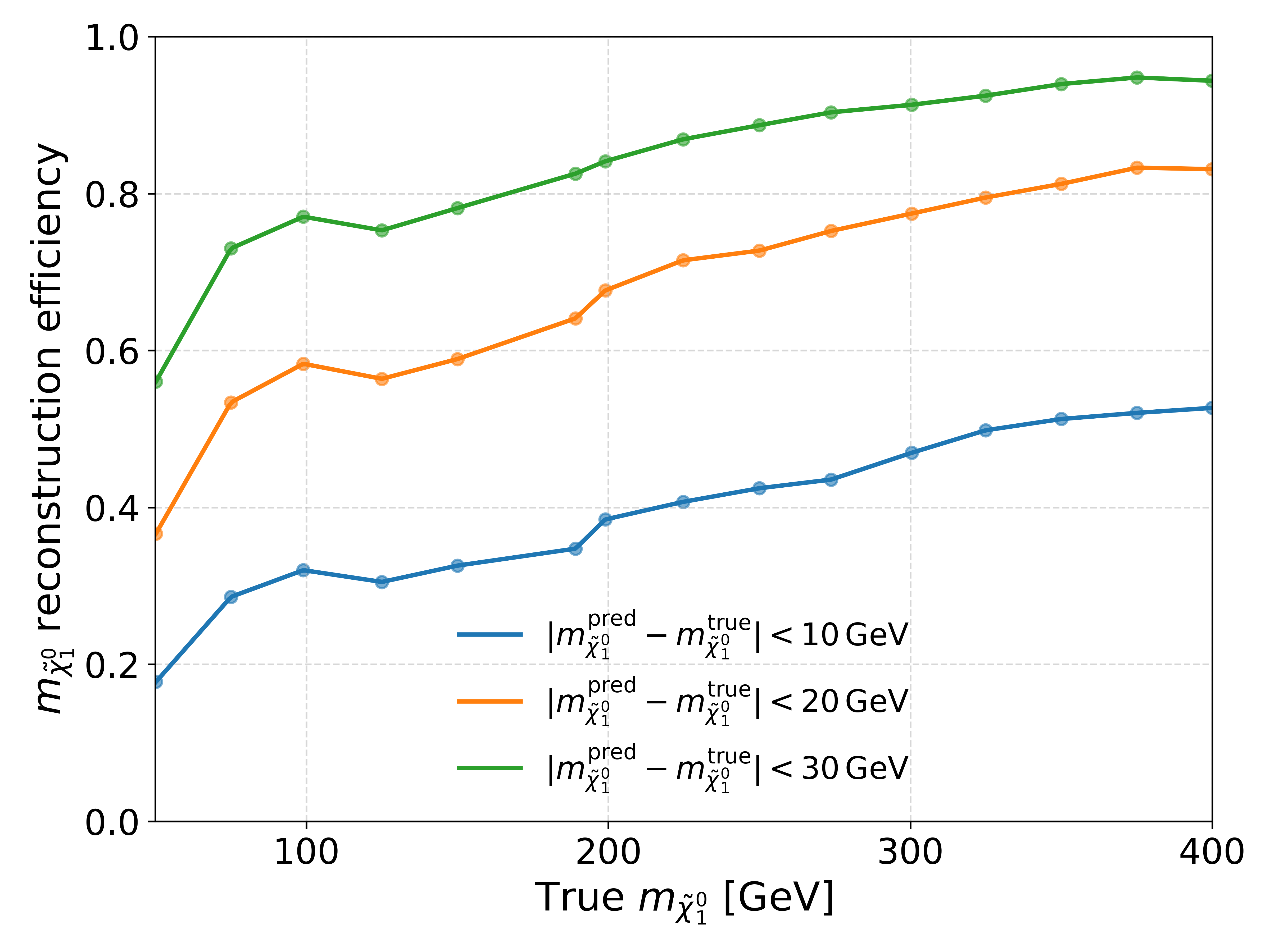}\hspace{0.3cm}\includegraphics[width=0.47\linewidth]{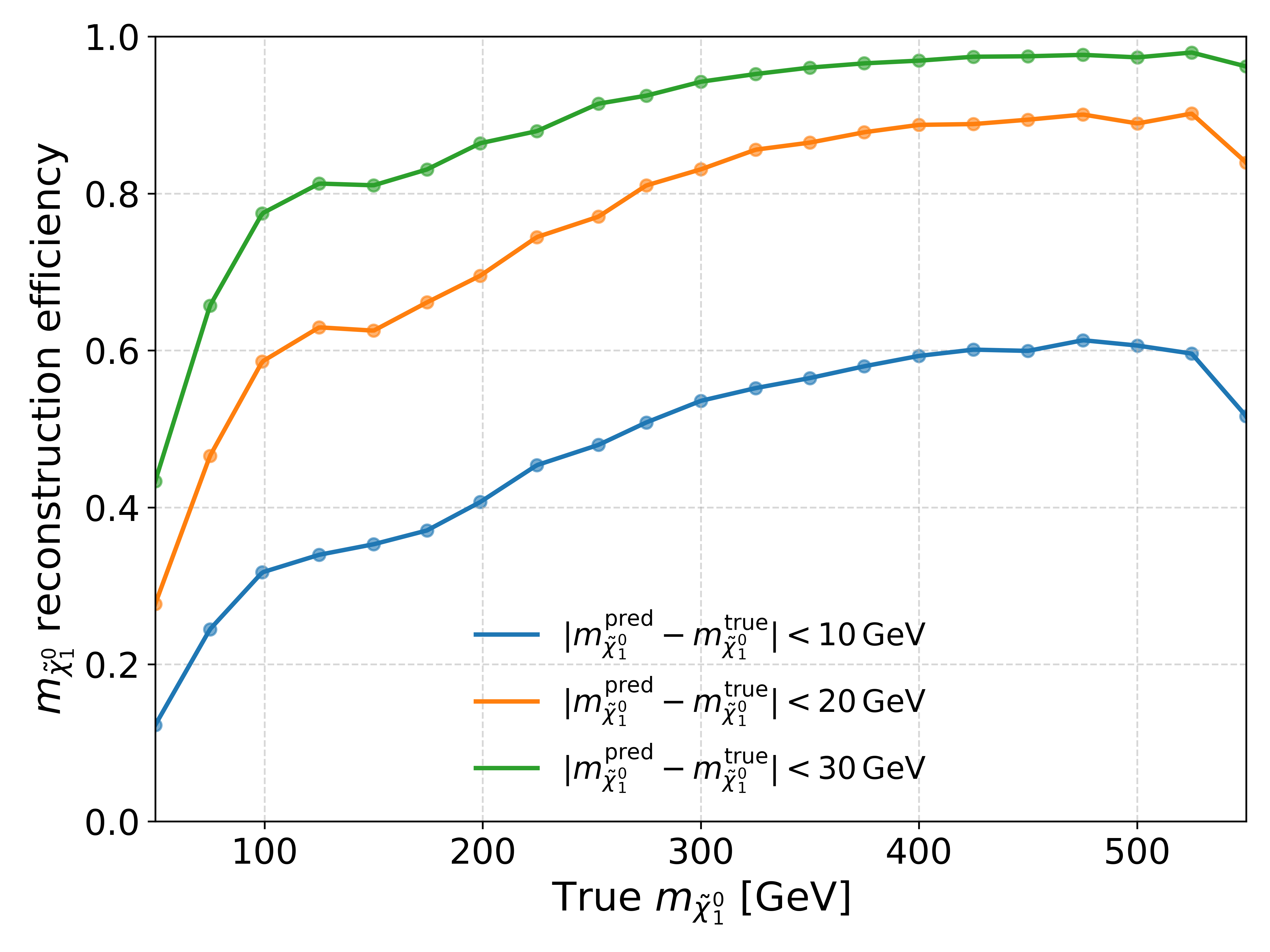}
\caption{The fraction of events where $|m_{\lspone}^{\mathrm{pred}} - m_{\lspone}^{\mathrm{true}}| < \Delta m$, are shown, for three tolerances, $\Delta m = 10,\,20$ and 30~GeV. Here, the network is trained on events with a fixed $m_{\lsptwo} = 1500~$GeV~(left panel) and 2000~GeV~(right panel), but with $m_{\lspone}$ varying between 50 and 550~GeV, with a 25~GeV step size. The network is evaluated on events generated with $m_{\lspone}$ varying over the same range.}
    \label{fig:reco_eff_nlsp_lle_600GeV}
\end{figure}

For training purposes, we use the following detector-level observables,
\begin{align}
    \{p_x,p_y,p_z,E\}_{\ell_i}, \cancel{p}_x, \cancel{p}_y, \Delta R(\ell_i,\ell_j),
    \Delta \phi (\ell_i,\cancel{E}_T) 
\end{align}
with $i,j=1~\text{to}~4$. Here, we put the four-momentum information of the leptons in one token and other variables into a separate token. We use a similar procedure and the same set of hyperparameters for our training as the UDD benchmark scenario. To show the performance of our ML network, we set two values for $\mlsptwo$, with $\mlsptwo=1.5$ TeV and $\mlsptwo=2$ TeV. We vary $\mlspone$ from 50 to 550 GeV with a 25 GeV increment in the step size at each step. The fraction of events with $|m_{\tilde\chi_1^0}^{\text{pred}}-m_{\tilde\chi_1^0}^{\text{true}}|<\Delta m$ with $\Delta m=10,20$ and $30$ GeV are shown in \ref{fig:reco_eff_nlsp_lle_600GeV}. For $\Delta m = 30~$GeV, we achieve $\gtrsim 80\%$ $\mlspone$ reconstruction efficiency for $m_{\lspone}$ in the range of $150~\textrm{GeV} \lesssim  m_{\lspone} \lesssim 550~\rm{GeV}$, for $m_{\lsptwo}=1.5~$ and 2 TeV. However, we note that the reconstruction efficiency falls near the edges. These results indicate the generalizability of our network for reconstructing masses in different processes.


\begin{thebibliography}{10}

\bibitem{SUSSKIND1984181}
L.~Susskind, ``The gauge hierarchy problem, technicolor, supersymmetry, and all
  that,''
  \href{http://dx.doi.org/https://doi.org/10.1016/0370-1573(84)90208-4}{{\em
  Physics Reports} {\bfseries 104} no.~2, (1984) 181--193}.
  \url{https://www.sciencedirect.com/science/article/pii/0370157384902084}.

\bibitem{PhysRevD.14.1667}
E.~Gildener, ``Gauge-symmetry hierarchies,''
  \href{http://dx.doi.org/10.1103/PhysRevD.14.1667}{{\em Phys. Rev. D}
  {\bfseries 14} (Sep, 1976) 1667--1672}.
  \url{https://link.aps.org/doi/10.1103/PhysRevD.14.1667}.

\bibitem{KamLAND:2013rgu}
{\bfseries KamLAND} Collaboration, A.~Gando {\em et~al.}, ``{Reactor On-Off
  Antineutrino Measurement with KamLAND},''
  \href{http://dx.doi.org/10.1103/PhysRevD.88.033001}{{\em Phys. Rev. D}
  {\bfseries 88} no.~3, (2013) 033001},
  \href{http://arxiv.org/abs/1303.4667}{{\ttfamily arXiv:1303.4667 [hep-ex]}}.

\bibitem{Borexino:2013zhu}
{\bfseries Borexino} Collaboration, G.~Bellini {\em et~al.}, ``{Final results
  of Borexino Phase-I on low energy solar neutrino spectroscopy},''
  \href{http://dx.doi.org/10.1103/PhysRevD.89.112007}{{\em Phys. Rev. D}
  {\bfseries 89} no.~11, (2014) 112007},
  \href{http://arxiv.org/abs/1308.0443}{{\ttfamily arXiv:1308.0443 [hep-ex]}}.

\bibitem{RENO:2018dro}
{\bfseries RENO} Collaboration, G.~Bak {\em et~al.}, ``{Measurement of Reactor
  Antineutrino Oscillation Amplitude and Frequency at RENO},''
  \href{http://dx.doi.org/10.1103/PhysRevLett.121.201801}{{\em Phys. Rev.
  Lett.} {\bfseries 121} no.~20, (2018) 201801},
  \href{http://arxiv.org/abs/1806.00248}{{\ttfamily arXiv:1806.00248
  [hep-ex]}}.

\bibitem{T2K:2018rhz}
{\bfseries T2K} Collaboration, K.~Abe {\em et~al.}, ``{Search for CP Violation
  in Neutrino and Antineutrino Oscillations by the T2K Experiment with
  $2.2\times10^{21}$ Protons on Target},''
  \href{http://dx.doi.org/10.1103/PhysRevLett.121.171802}{{\em Phys. Rev.
  Lett.} {\bfseries 121} no.~17, (2018) 171802},
  \href{http://arxiv.org/abs/1807.07891}{{\ttfamily arXiv:1807.07891
  [hep-ex]}}.

\bibitem{DayaBay:2018yms}
{\bfseries Daya Bay} Collaboration, D.~Adey {\em et~al.}, ``{Measurement of the
  Electron Antineutrino Oscillation with 1958 Days of Operation at Daya Bay},''
  \href{http://dx.doi.org/10.1103/PhysRevLett.121.241805}{{\em Phys. Rev.
  Lett.} {\bfseries 121} no.~24, (2018) 241805},
  \href{http://arxiv.org/abs/1809.02261}{{\ttfamily arXiv:1809.02261
  [hep-ex]}}.

\bibitem{Super-Kamiokande:2019gzr}
{\bfseries Super-Kamiokande} Collaboration, M.~Jiang {\em et~al.},
  ``{Atmospheric Neutrino Oscillation Analysis with Improved Event
  Reconstruction in Super-Kamiokande IV},''
  \href{http://dx.doi.org/10.1093/ptep/ptz015}{{\em PTEP} {\bfseries 2019}
  no.~5, (2019) 053F01}, \href{http://arxiv.org/abs/1901.03230}{{\ttfamily
  arXiv:1901.03230 [hep-ex]}}.

\bibitem{NOvA:2019cyt}
{\bfseries NOvA} Collaboration, M.~A. Acero {\em et~al.}, ``{First Measurement
  of Neutrino Oscillation Parameters using Neutrinos and Antineutrinos by
  NOvA},'' \href{http://dx.doi.org/10.1103/PhysRevLett.123.151803}{{\em Phys.
  Rev. Lett.} {\bfseries 123} no.~15, (2019) 151803},
  \href{http://arxiv.org/abs/1906.04907}{{\ttfamily arXiv:1906.04907
  [hep-ex]}}.

\bibitem{Zwicky:1933gu}
F.~Zwicky, ``{Die Rotverschiebung von extragalaktischen Nebeln},''
  \href{http://dx.doi.org/10.1007/s10714-008-0707-4}{{\em Helv. Phys. Acta}
  {\bfseries 6} (1933) 110--127}.

\bibitem{1937ApJ....86..217Z}
F.~{Zwicky}, ``{On the Masses of Nebulae and of Clusters of Nebulae},''
  \href{http://dx.doi.org/10.1086/143864}{{\em apj} {\bfseries 86} (Oct., 1937)
  217}.

\bibitem{Jungman:1995df}
G.~Jungman, M.~Kamionkowski, and K.~Griest, ``{Supersymmetric dark matter},''
  \href{http://dx.doi.org/10.1016/0370-1573(95)00058-5}{{\em Phys. Rept.}
  {\bfseries 267} (1996) 195--373},
  \href{http://arxiv.org/abs/hep-ph/9506380}{{\ttfamily arXiv:hep-ph/9506380}}.

\bibitem{Sofue:2000jx}
Y.~Sofue and V.~Rubin, ``{Rotation curves of spiral galaxies},''
  \href{http://dx.doi.org/10.1146/annurev.astro.39.1.137}{{\em Ann. Rev.
  Astron. Astrophys.} {\bfseries 39} (2001) 137--174},
  \href{http://arxiv.org/abs/astro-ph/0010594}{{\ttfamily
  arXiv:astro-ph/0010594}}.

\bibitem{deOliveira:2017pjk}
L.~de~Oliveira, M.~Paganini, and B.~Nachman, ``{Learning Particle Physics by
  Example: Location-Aware Generative Adversarial Networks for Physics
  Synthesis},'' \href{http://dx.doi.org/10.1007/s41781-017-0004-6}{{\em Comput.
  Softw. Big Sci.} {\bfseries 1} no.~1, (2017) 4},
  \href{http://arxiv.org/abs/1701.05927}{{\ttfamily arXiv:1701.05927
  [stat.ML]}}.

\bibitem{Paganini:2017hrr}
M.~Paganini, L.~de~Oliveira, and B.~Nachman, ``{Accelerating Science with
  Generative Adversarial Networks: An Application to 3D Particle Showers in
  Multilayer Calorimeters},''
  \href{http://dx.doi.org/10.1103/PhysRevLett.120.042003}{{\em Phys. Rev.
  Lett.} {\bfseries 120} no.~4, (2018) 042003},
  \href{http://arxiv.org/abs/1705.02355}{{\ttfamily arXiv:1705.02355
  [hep-ex]}}.

\bibitem{Paganini:2017dwg}
M.~Paganini, L.~de~Oliveira, and B.~Nachman, ``{CaloGAN : Simulating 3D high
  energy particle showers in multilayer electromagnetic calorimeters with
  generative adversarial networks},''
  \href{http://dx.doi.org/10.1103/PhysRevD.97.014021}{{\em Phys. Rev. D}
  {\bfseries 97} no.~1, (2018) 014021},
  \href{http://arxiv.org/abs/1712.10321}{{\ttfamily arXiv:1712.10321
  [hep-ex]}}.

\bibitem{Datta:2018mwd}
K.~Datta, D.~Kar, and D.~Roy, ``{Unfolding with Generative Adversarial
  Networks},'' \href{http://arxiv.org/abs/1806.00433}{{\ttfamily
  arXiv:1806.00433 [physics.data-an]}}.

\bibitem{Butter:2019cae}
A.~Butter, T.~Plehn, and R.~Winterhalder, ``{How to GAN LHC Events},''
  \href{http://dx.doi.org/10.21468/SciPostPhys.7.6.075}{{\em SciPost Phys.}
  {\bfseries 7} no.~6, (2019) 075},
  \href{http://arxiv.org/abs/1907.03764}{{\ttfamily arXiv:1907.03764
  [hep-ph]}}.

\bibitem{Bellagente:2019uyp}
M.~Bellagente, A.~Butter, G.~Kasieczka, T.~Plehn, and R.~Winterhalder, ``{How
  to GAN away Detector Effects},''
  \href{http://dx.doi.org/10.21468/SciPostPhys.8.4.070}{{\em SciPost Phys.}
  {\bfseries 8} no.~4, (2020) 070},
  \href{http://arxiv.org/abs/1912.00477}{{\ttfamily arXiv:1912.00477
  [hep-ph]}}.

\bibitem{Belayneh:2019vyx}
D.~Belayneh {\em et~al.}, ``{Calorimetry with deep learning: particle
  simulation and reconstruction for collider physics},''
  \href{http://dx.doi.org/10.1140/epjc/s10052-020-8251-9}{{\em Eur. Phys. J. C}
  {\bfseries 80} no.~7, (2020) 688},
  \href{http://arxiv.org/abs/1912.06794}{{\ttfamily arXiv:1912.06794
  [physics.ins-det]}}.

\bibitem{Gao:2020zvv}
C.~Gao, S.~H{\"o}che, J.~Isaacson, C.~Krause, and H.~Schulz, ``{Event
  Generation with Normalizing Flows},''
  \href{http://dx.doi.org/10.1103/PhysRevD.101.076002}{{\em Phys. Rev. D}
  {\bfseries 101} no.~7, (2020) 076002},
  \href{http://arxiv.org/abs/2001.10028}{{\ttfamily arXiv:2001.10028
  [hep-ph]}}.

\bibitem{Carminati_2020}
F.~Carminati, G.~Khattak, V.~Loncar, T.~Q. Nguyen, M.~Pierini, R.~Brito
  Da~Rocha, K.~Samaras-Tsakiris, S.~Vallecorsa, and J.-R. Vlimant, ``Generative
  adversarial networks for fast simulation,''
  \href{http://dx.doi.org/10.1088/1742-6596/1525/1/012064}{{\em Journal of
  Physics: Conference Series} {\bfseries 1525} no.~1, (Apr, 2020) 012064}.
  \url{https://dx.doi.org/10.1088/1742-6596/1525/1/012064}.

\bibitem{Bellagente:2020piv}
M.~Bellagente, A.~Butter, G.~Kasieczka, T.~Plehn, A.~Rousselot,
  R.~Winterhalder, L.~Ardizzone, and U.~K\"othe, ``{Invertible Networks or
  Partons to Detector and Back Again},''
  \href{http://dx.doi.org/10.21468/SciPostPhys.9.5.074}{{\em SciPost Phys.}
  {\bfseries 9} (2020) 074}, \href{http://arxiv.org/abs/2006.06685}{{\ttfamily
  arXiv:2006.06685 [hep-ph]}}.

\bibitem{Butter:2020tvl}
A.~Butter and T.~Plehn, ``{Generative Networks for LHC events},''
  \href{http://arxiv.org/abs/2008.08558}{{\ttfamily arXiv:2008.08558
  [hep-ph]}}.

\bibitem{Krause:2021ilc}
C.~Krause and D.~Shih, ``{Fast and accurate simulations of calorimeter showers
  with normalizing flows},''
  \href{http://dx.doi.org/10.1103/PhysRevD.107.113003}{{\em Phys. Rev. D}
  {\bfseries 107} no.~11, (2023) 113003},
  \href{http://arxiv.org/abs/2106.05285}{{\ttfamily arXiv:2106.05285
  [physics.ins-det]}}.

\bibitem{Kansal:2021cqp}
R.~Kansal, J.~Duarte, H.~Su, B.~Orzari, T.~Tomei, M.~Pierini, M.~Touranakou,
  J.-R. Vlimant, and D.~Gunopulos, ``{Particle Cloud Generation with Message
  Passing Generative Adversarial Networks},'' in {\em {35th Conference on
  Neural Information Processing Systems}}.
\newblock 6, 2021.
\newblock \href{http://arxiv.org/abs/2106.11535}{{\ttfamily arXiv:2106.11535
  [cs.LG]}}.

\bibitem{Bieringer:2022cbs}
S.~Bieringer, A.~Butter, S.~Diefenbacher, E.~Eren, F.~Gaede, D.~Hundhausen,
  G.~Kasieczka, B.~Nachman, T.~Plehn, and M.~Trabs, ``{Calomplification
  {\textemdash} the power of generative calorimeter models},''
  \href{http://dx.doi.org/10.1088/1748-0221/17/09/P09028}{{\em JINST}
  {\bfseries 17} no.~09, (2022) P09028},
  \href{http://arxiv.org/abs/2202.07352}{{\ttfamily arXiv:2202.07352
  [hep-ph]}}.

\bibitem{Touranakou:2022qrp}
M.~Touranakou, N.~Chernyavskaya, J.~Duarte, D.~Gunopulos, R.~Kansal, B.~Orzari,
  M.~Pierini, T.~Tomei, and J.-R. Vlimant, ``{Particle-based fast jet
  simulation at the LHC with variational autoencoders},''
  \href{http://dx.doi.org/10.1088/2632-2153/ac7c56}{{\em Mach. Learn. Sci.
  Tech.} {\bfseries 3} no.~3, (2022) 035003},
  \href{http://arxiv.org/abs/2203.00520}{{\ttfamily arXiv:2203.00520
  [physics.comp-ph]}}.

\bibitem{Butter:2022vkj}
A.~Butter, T.~Heimel, T.~Martini, S.~Peitzsch, and T.~Plehn, ``{Two invertible
  networks for the matrix element method},''
  \href{http://dx.doi.org/10.21468/SciPostPhys.15.3.094}{{\em SciPost Phys.}
  {\bfseries 15} no.~3, (2023) 094},
  \href{http://arxiv.org/abs/2210.00019}{{\ttfamily arXiv:2210.00019
  [hep-ph]}}.

\bibitem{ATLAS:2022jhk}
{\bfseries ATLAS} Collaboration, G.~Aad {\em et~al.}, ``{Deep Generative Models
  for Fast Photon Shower Simulation in ATLAS},''
  \href{http://dx.doi.org/10.1007/s41781-023-00106-9}{{\em Comput. Softw. Big
  Sci.} {\bfseries 8} no.~1, (2024) 7},
  \href{http://arxiv.org/abs/2210.06204}{{\ttfamily arXiv:2210.06204
  [hep-ex]}}.

\bibitem{Backes:2022sph}
M.~Backes, A.~Butter, M.~Dunford, and B.~Malaescu, ``{An unfolding method based
  on conditional invertible neural networks (cINN) using iterative training},''
  \href{http://dx.doi.org/10.21468/scipostphyscore.7.1.007}{{\em SciPost Phys.
  Core} {\bfseries 7} no.~1, (2024) 007},
  \href{http://arxiv.org/abs/2212.08674}{{\ttfamily arXiv:2212.08674
  [hep-ph]}}.

\bibitem{Mikuni:2023dvk}
V.~Mikuni, B.~Nachman, and M.~Pettee, ``{Fast point cloud generation with
  diffusion models in high energy physics},''
  \href{http://dx.doi.org/10.1103/PhysRevD.108.036025}{{\em Phys. Rev. D}
  {\bfseries 108} no.~3, (2023) 036025},
  \href{http://arxiv.org/abs/2304.01266}{{\ttfamily arXiv:2304.01266
  [hep-ph]}}.

\bibitem{Ackerschott:2023nax}
J.~Ackerschott, R.~K. Barman, D.~Gon\c{c}alves, T.~Heimel, and T.~Plehn,
  ``{Returning CP-Observables to The Frames They Belong},''
  \href{http://arxiv.org/abs/2308.00027}{{\ttfamily arXiv:2308.00027
  [hep-ph]}}.

\bibitem{Butter:2023ira}
A.~Butter, T.~Jezo, M.~Klasen, M.~Kuschick, S.~Palacios~Schweitzer, and
  T.~Plehn, ``{Kicking it off(-shell) with direct diffusion},''
  \href{http://dx.doi.org/10.21468/SciPostPhysCore.7.3.064}{{\em SciPost Phys.
  Core} {\bfseries 7} no.~3, (2024) 064},
  \href{http://arxiv.org/abs/2311.17175}{{\ttfamily arXiv:2311.17175
  [hep-ph]}}.

\bibitem{Butter:2023llt}
A.~Butter, ``{Normalizing Flows for LHC Theory},''
  \href{http://dx.doi.org/10.1088/1742-6596/2438/1/012004}{{\em J. Phys. Conf.
  Ser.} {\bfseries 2438} no.~1, (2023) 012004}.

\bibitem{Huetsch:2024quz}
N.~Huetsch {\em et~al.}, ``{The landscape of unfolding with machine
  learning},'' \href{http://dx.doi.org/10.21468/SciPostPhys.18.2.070}{{\em
  SciPost Phys.} {\bfseries 18} no.~2, (2025) 070},
  \href{http://arxiv.org/abs/2404.18807}{{\ttfamily arXiv:2404.18807
  [hep-ph]}}.

\bibitem{Butter:2024vbx}
A.~Butter, S.~Diefenbacher, N.~Huetsch, V.~Mikuni, B.~Nachman,
  S.~Palacios~Schweitzer, and T.~Plehn, ``{Generative Unfolding with
  Distribution Mapping},''
  \href{http://dx.doi.org/10.21468/SciPostPhys.18.6.200}{{\em SciPost Phys.}
  {\bfseries 18} (2025) 200}, \href{http://arxiv.org/abs/2411.02495}{{\ttfamily
  arXiv:2411.02495 [hep-ph]}}.

\bibitem{Abasov:2025ntj}
E.~Abasov, L.~Dudko, E.~Iudin, A.~Markina, P.~Volkov, G.~Vorotnikov,
  M.~Perfilov, and A.~Zaborenko, ``{Reconstruction of angular correlations in
  the associated top quark and the dark matter mediator production},''
  \href{http://arxiv.org/abs/2504.14303}{{\ttfamily arXiv:2504.14303
  [hep-ph]}}.

\bibitem{Chatterjee:2025gej}
A.~Chatterjee, A.~Choudhury, S.~Mitra, A.~Mondal, and S.~Mondal, ``{Exploring
  the BSM parameter space with Neural Network aided Simulation-Based
  Inference},'' \href{http://arxiv.org/abs/2502.11928}{{\ttfamily
  arXiv:2502.11928 [hep-ph]}}.

\bibitem{goodfellow2014generative}
I.~J. Goodfellow, J.~Pouget-Abadie, M.~Mirza, B.~Xu, D.~Warde-Farley, S.~Ozair,
  A.~Courville, and Y.~Bengio, ``Generative adversarial networks,'' 2014.

\bibitem{rezende2014}
D.~J. Rezende, S.~Mohamed, and D.~Wierstra, ``Stochastic backpropagation and
  approximate inference in deep generative models,'' 2014.
\newblock \url{https://arxiv.org/abs/1401.4082}.

\bibitem{kingma2022}
D.~P. Kingma and M.~Welling, ``Auto-encoding variational bayes,'' 2022.
\newblock \url{https://arxiv.org/abs/1312.6114}.

\bibitem{dinh2015nice}
L.~Dinh, D.~Krueger, and Y.~Bengio, ``Nice: Non-linear independent components
  estimation,'' 2015.

\bibitem{sohldickstein2015deep}
J.~Sohl-Dickstein, E.~A. Weiss, N.~Maheswaranathan, and S.~Ganguli, ``Deep
  unsupervised learning using nonequilibrium thermodynamics,'' 2015.

\bibitem{Dreiner:2023bvs}
H.~K. Dreiner, Y.~S. Koay, D.~K{\"o}hler, V.~M. Lozano, J.~Montejo~Berlingen,
  S.~Nangia, and N.~Strobbe, ``{The ABC of RPV: classification of R-parity
  violating signatures at the LHC for small couplings},''
  \href{http://dx.doi.org/10.1007/JHEP07(2023)215}{{\em JHEP} {\bfseries 07}
  (2023) 215}, \href{http://arxiv.org/abs/2306.07317}{{\ttfamily
  arXiv:2306.07317 [hep-ph]}}.

\bibitem{Choudhury:2023lbp}
A.~Choudhury, S.~Mitra, A.~Mondal, and S.~Mondal, ``{Bilinear R-parity
  violating supersymmetry under the light of neutrino oscillation, Higgs and
  flavor data},'' \href{http://dx.doi.org/10.1007/JHEP02(2024)004}{{\em JHEP}
  {\bfseries 02} (2024) 004}, \href{http://arxiv.org/abs/2305.15211}{{\ttfamily
  arXiv:2305.15211 [hep-ph]}}.

\bibitem{Choudhury:2023eje}
A.~Choudhury, A.~Mondal, S.~Mondal, and S.~Sarkar, ``{Improving sensitivity of
  trilinear R-parity violating SUSY searches using machine learning at the
  LHC},'' \href{http://dx.doi.org/10.1103/PhysRevD.109.035001}{{\em Phys. Rev.
  D} {\bfseries 109} no.~3, (2024) 035001},
  \href{http://arxiv.org/abs/2308.02697}{{\ttfamily arXiv:2308.02697
  [hep-ph]}}.

\bibitem{Choudhury:2023yfg}
A.~Choudhury, A.~Mondal, S.~Mondal, and S.~Sarkar, ``{Slepton searches in the
  trilinear RPV SUSY scenarios at the HL-LHC and HE-LHC},''
  \href{http://arxiv.org/abs/2310.07532}{{\ttfamily arXiv:2310.07532
  [hep-ph]}}.

\bibitem{Choudhury:2024yxd}
A.~Choudhury, S.~Mitra, A.~Mondal, and S.~Mondal, ``{An MCMC analysis to probe
  trilinear RPV SUSY scenarios and possible LHC signatures},''
  \href{http://arxiv.org/abs/2411.08112}{{\ttfamily arXiv:2411.08112
  [hep-ph]}}.

\bibitem{Baruah:2024wrn}
R.~Baruah, A.~Choudhury, K.~Ghosh, S.~Mondal, and R.~Sahu, ``{Probing sub-TeV
  Higgsinos aided by a machine-learning-based top tagger in the context of
  trilinear R-parity violating SUSY},''
  \href{http://dx.doi.org/10.1103/PhysRevD.111.095004}{{\em Phys. Rev. D}
  {\bfseries 111} no.~9, (2025) 095004},
  \href{http://arxiv.org/abs/2412.11862}{{\ttfamily arXiv:2412.11862
  [hep-ph]}}.

\bibitem{Dreiner:2025kfd}
H.~K. Dreiner, M.~Hank, Y.~S. Koay, M.~Sch{\"u}rmann, R.~Sengupta, A.~Shah,
  N.~Strobbe, and E.~Thomson, ``{The ABC of RPV. Part II. Classification of
  R-parity violating signatures from UDD couplings and their coverage at the
  LHC},'' \href{http://dx.doi.org/10.1007/JHEP06(2025)258}{{\em JHEP}
  {\bfseries 06} (2025) 258}, \href{http://arxiv.org/abs/2503.03830}{{\ttfamily
  arXiv:2503.03830 [hep-ph]}}.

\bibitem{Dreiner:1997uz}
H.~K. Dreiner, ``{An Introduction to explicit R-parity violation},''
  \href{http://dx.doi.org/10.1142/9789814307505_0017}{{\em Adv. Ser. Direct.
  High Energy Phys.} {\bfseries 21} (2010) 565--583},
  \href{http://arxiv.org/abs/hep-ph/9707435}{{\ttfamily arXiv:hep-ph/9707435}}.

\bibitem{Martin:1997ns}
S.~P. Martin, ``{A Supersymmetry primer},''
  \href{http://dx.doi.org/10.1142/9789812839657_0001}{{\em Adv. Ser. Direct.
  High Energy Phys.} {\bfseries 18} (1998) 1--98},
  \href{http://arxiv.org/abs/hep-ph/9709356}{{\ttfamily arXiv:hep-ph/9709356}}.

\bibitem{Barbier:2004ez}
R.~Barbier {\em et~al.}, ``{R-parity violating supersymmetry},''
  \href{http://dx.doi.org/10.1016/j.physrep.2005.08.006}{{\em Phys. Rept.}
  {\bfseries 420} (2005) 1--202},
  \href{http://arxiv.org/abs/hep-ph/0406039}{{\ttfamily arXiv:hep-ph/0406039}}.

\bibitem{Choudhury:2024ggy}
A.~Choudhury, A.~Mondal, and S.~Mondal, ``{Status of R-parity violating
  SUSY},'' \href{http://arxiv.org/abs/2402.04040}{{\ttfamily arXiv:2402.04040
  [hep-ph]}}.

\bibitem{Barman:2020azo}
R.~K. Barman, B.~Bhattacherjee, I.~Chakraborty, A.~Choudhury, and N.~Khan,
  ``{Electroweakino searches at the HL-LHC in the baryon number violating
  MSSM},'' \href{http://dx.doi.org/10.1103/PhysRevD.103.015003}{{\em Phys. Rev.
  D} {\bfseries 103} no.~1, (2021) 015003},
  \href{http://arxiv.org/abs/2003.10920}{{\ttfamily arXiv:2003.10920
  [hep-ph]}}.

\bibitem{Sjostrand:2006za}
T.~Sjostrand, S.~Mrenna, and P.~Z. Skands, ``{PYTHIA 6.4 Physics and Manual},''
  \href{http://dx.doi.org/10.1088/1126-6708/2006/05/026}{{\em JHEP} {\bfseries
  05} (2006) 026}, \href{http://arxiv.org/abs/hep-ph/0603175}{{\ttfamily
  arXiv:hep-ph/0603175}}.

\bibitem{deFavereau:2013fsa}
{\bfseries DELPHES 3} Collaboration, J.~de~Favereau, C.~Delaere, P.~Demin,
  A.~Giammanco, V.~Lema\^\i{}tre, A.~Mertens, and M.~Selvaggi, ``{DELPHES 3, A
  modular framework for fast simulation of a generic collider experiment},''
  \href{http://dx.doi.org/10.1007/JHEP02(2014)057}{{\em JHEP} {\bfseries 02}
  (2014) 057}, \href{http://arxiv.org/abs/1307.6346}{{\ttfamily arXiv:1307.6346
  [hep-ex]}}.

\bibitem{vaswani2017attention}
A.~Vaswani, N.~Shazeer, N.~Parmar, J.~Uszkoreit, L.~Jones, A.~N. Gomez,
  {\L}.~Kaiser, and I.~Polosukhin, ``Attention is all you need,'' {\em Advances
  in neural information processing systems} {\bfseries 30} (2017) .

\bibitem{hendrycks2023gaussianerrorlinearunits}
D.~Hendrycks and K.~Gimpel, ``Gaussian error linear units (gelus),'' 2023.
\newblock \url{https://arxiv.org/abs/1606.08415}.

\bibitem{ho2020denoising}
J.~Ho, A.~Jain, and P.~Abbeel, ``Denoising diffusion probabilistic models,''
  2020.

\bibitem{paszke2019pytorchimperativestylehighperformance}
A.~Paszke, S.~Gross, F.~Massa, A.~Lerer, J.~Bradbury, G.~Chanan, T.~Killeen,
  Z.~Lin, N.~Gimelshein, L.~Antiga, A.~Desmaison, A.~Köpf, E.~Yang, Z.~DeVito,
  M.~Raison, A.~Tejani, S.~Chilamkurthy, B.~Steiner, L.~Fang, J.~Bai, and
  S.~Chintala, ``Pytorch: An imperative style, high-performance deep learning
  library,'' 2019.
\newblock \url{https://arxiv.org/abs/1912.01703}.

\bibitem{ATLAS:2021yyr}
{\bfseries ATLAS} Collaboration, G.~Aad {\em et~al.}, ``{Search for
  supersymmetry in events with four or more charged leptons in 139
  $\rm{fb}^{−1}$ of $ \sqrt{s} $ = 13 TeV pp collisions with the ATLAS
  detector},'' \href{http://dx.doi.org/10.1007/JHEP07(2021)167}{{\em JHEP}
  {\bfseries 07} (2021) 167}, \href{http://arxiv.org/abs/2103.11684}{{\ttfamily
  arXiv:2103.11684 [hep-ex]}}.

\end{thebibliography}
\providecommand{\href}[2]{#2}\begingroup\raggedright\endgroup
\end{document}